\documentclass[nohypdvips,review,hidelinks,onefignum,onetabnum,bookmarks=true]{siamart250211}

\usepackage{fontspec}
\usepackage{lmodern}

\usepackage{polyglossia}
\setdefaultlanguage{english}
\usepackage{amsmath}
\usepackage{amssymb}
\usepackage{float}

\usepackage{graphicx}
\usepackage{geometry}
\usepackage{pdfpages}
\usepackage{pgfgantt}

\usepackage{siunitx}
\usepackage{wrapfig}
\usepackage{xcolor}
\usepackage{xspace}

\usepackage{enumitem}
\usepackage{listings}
\usepackage{subcaption}
\usepackage{xfrac}
\usepackage{appendix}
\usepackage{physics}
\usepackage{bm}
\usepackage{siunitx}

\usepackage{algorithmic}
\usepackage{mathtools}
\usepackage{physics}

\usetikzlibrary{positioning,arrows.meta,calc,fit,backgrounds}

\newcommand{\stackbehind}[2]{
  \begin{scope}[on background layer]
    \foreach \d in {3.2mm,1.6mm}{
      \draw[thick, rounded corners=2pt, fill=#2]
        ($(#1.south west)+(\d,\d)$) rectangle ($(#1.north east)+(\d,\d)$);
    }
  \end{scope}
}
\title{Hierarchical Bayesian Calibration with Bayesian Committee Machine
  \thanks{\funding{This work was funded by the Swiss National Science Foundation.}}}

\author{Sebastian Heinekamp\thanks{Paul Scherrer Institut, Villigen,
    Switzerland, and ETH Z\"urich, Z\"urich, Switzerland.}
  \and David M. Higdon\thanks{Virginia Tech, Blacksburg, VA, USA.}
  \and Andreas Adelmann\thanks{Paul Scherrer Institut, Villigen, Switzerland
    (\email{andreas.adelmann@psi.ch}). Corresponding author.}}

\date{\today} 
\headers{Hierarchical Bayesian Calibration with BCM}{S. Heinekamp, D. M. Higdon, and A. Adelmann}

\hypersetup{
    unicode=false,          
    pdftoolbar=true,        
    pdfmenubar=true,        
    pdffitwindow=true,      
    pdftitle={Bayesian Committee Machine in Hierarchical Bayesian Calibration},    
    pdfauthor={Sebastian Heinekamp},     
    pdfsubject={Subject},   
    pdfnewwindow=true,      
    pdfkeywords={keywords}, 
    colorlinks=true,        
    linkcolor=black,        
    citecolor=blue,         
    filecolor=blue,         
    urlcolor=blue           
}

\newcommand{\GP}{GP\xspace}
\newcommand{\Eq}{Equation\xspace}
\newcommand{\BCM}{BCM\xspace}

\newcommand{\BC}{BayesCal\xspace}
\newcommand{\MBC}{committee member\xspace}
\newcommand{\MBCS}{committee members\xspace}

\begin{document}
\nolinenumbers 
\maketitle

\begin{abstract}
Calibrating computational models to experimental data is a core task in applied statistics, especially in scientific domains, where physical experiments are costly and simulations play a central role in design and inference. Motivated by uncertainty quantification challenges in particle accelerator experiments, we develop and evaluate a Hierarchical Bayesian Calibration framework.
In contrast to standard Bayesian calibration, certain inputs—such as beam injection amplitude—must be estimated separately for each experiment. We adopt the Kennedy–O’Hagan formulation and extend it with a hierarchical prior structure to model the distribution of experiment-specific calibration parameters, thus borrowing strength and improving generalisation across repeated experiments. A key methodological challenge arises from the need to evaluate a large number of forward simulations, which renders conventional Markov chain Monte Carlo (MCMC) approaches computationally prohibitive. To address this, we leverage the Bayesian Committee Machine (BCM) as a scalable modelling strategy for Gaussian Process emulators. The BCM provides a principled divide-and-conquer approach, enabling parallel inference and reducing computational cost without requiring problem-specific tuning of the emulator approximation.
Posterior sampling is performed using the No-U-Turn Sampler, supported by automatic differentiation in Julia, which removes the need for analytic gradient derivation and facilitates flexible model specification.\ We assess the proposed framework using established benchmark problems and simulated data from the Argonne Wakefield Accelerator. The results demonstrate substantial computational savings and robust calibration performance, highlighting the applicability of the method to large-scale scientific modelling problems.
\end{abstract}



\section{Introduction}
\label{intro}
When simulations of machines or experiments are used to interpret real-world observations and enhance predictive accuracy, it becomes essential to calibrate simulation parameters using available measurements. This challenge arises across diverse scientific and engineering domains where complex simulations guide experimental design and operation. Bayesian Calibration addresses this requirement.

The calibration setup presented is geared towards 
particle accelerators. Due to the complexity
of these machines, simulation codes are imperative in the design, operation, and optimisation of such complex scientific instruments. 
A good calibration of the computational model parameters can improve and simplify this endeavour. In addition, accurately characterising posterior uncertainty gives
more realistic expectations for all future runs.
This affects the optimisation towards experimental goals and improves the uncertainty quantification. 
Ideally, the calibration is fast enough to be updated during an experiment and can provide information to the operator. This also means that the calibration runtime is a core constraint. 

To accurately capture calibration parameters drifting between experimental runs, a Hierarchical Bayesian Calibration is necessary.
We augment the Hierarchical Bayesian Calibration framework with the Bayesian Committee Machine, facilitating efficient utilisation of large data sets while substantially reducing computational runtime. This combination is particularly valuable for applications that require repeated, fast calibration as operational conditions evolve.

The example application at the Argonne Wakefield Accelerator Beam Test Facility (AWA) is the calibration of electron gun parameters from data recorded at the first monitor. Despite the short beamline considered, similar problems are ubiquitous in longer sections of this or other particle accelerator facilities worldwide.

Bayesian inference is very good at the efficient use of very few data points~\cite{tippingBayesianInferenceIntroduction2004}.
Hence, it is ideal for our particle accelerator application for which only a few physical measurements can be made in any given experiment.

Kennedy and O'Hagan~\cite{kennedyBayesianCalibrationComputer2001} introduced Bayesian Calibration (\BC) which had a large impact, and numerous improvements and additions have been made to it. To give a rough historical lineup, there is the MCMC-based implementation of Higdon et al.~\cite{higdonCombiningFieldData2004}, the work of Bayarri on highly irregular data~\cite{bayarriComputerModelValidation2007,bayarriPredictingVehicleCrashworthiness2009}, the extension of Higdon et al. to high-dimensional output~\cite{higdonComputerModelCalibration2008}, Goldstein and Rougier~\cite{goldsteinReifiedBayesianModelling2009} extend the model-inadequacy to account for multiple models, 

Qian \& Wu's adaptation to hierarchical models~\cite{qianBayesianHierarchicalModeling2008}, the derivation of convergence guarantees by Tuo \& Wu~\cite{tuoEfficientCalibrationImperfect2015} and the additional orthogonality of the code inadequacy $\delta(x)$ to the simulation code requirements $\eta(x,\theta)$ by Plumlee, which reduces the predicted uncertainty~\cite{plumleeBayesianCalibrationInexact2017}.

The ordinary Bayesian Calibration approach assumes a common set of calibration parameters for all of the physical experiments or measurements being considered. 
However, this does not suit our motivating particle accelerator application. Here, physical parameters change slightly with each experimental run of the accelerator. Hence, there is a need to repeatedly calibrate model parameters while leveraging prior calibrations since parameters will only change slightly from one run to the next.
Therefore, we add hierarchically linked  parameters~\cite{nagelUnifiedFrameworkMultilevel2016} to allow different parameter values for the various experiments being considered in the analysis. 
With this hierarchical treatment of parameters, we can more realistically model a collection of experiments, producing more reliable results. 
 However, this increase in the dimensionality of the parameter vector will add to the computational burden of posterior exploration via MCMC. 

Although real-world experiments are still expensive or infeasible, advances in computational power and numerical methods mean that it is feasible to gather a lot more simulation data. While Bayesian Statistics excels on small data sets, the use of MCMC in common Gaussian Process based (\GP-based) formulations involves demanding matrix computations and is typically impractical when working with thousands of data points or more without some type of approximation. Since the \GP is evaluated in each Monte Carlo step, the evaluation time of the \GP governs the runtime of the entire method. 

A common approach to reduce the evaluation time of the \GP posterior density for larger data sets is the approach of optimal subsets. These approaches reduce the data set to the data points that contain the most information~\cite{rasmussenGaussianProcessesMachine2005}. An advanced version of this interpolates to auxiliary data points to achieve a good approximation of the original data set as a whole. However, these approaches assume that the underlying function approximated by the \GP moves slowly enough in its inputs to be approximated well with a limited number of support points. Should this number of support points not be enough, information from the data set is lost. On the computational side, the selection of the subset 

has its own computational cost, growing with the size of the data set, depending on the selection strategy. 

A prominent instance of this approach is the Variational Sparse Gaussian Process~\cite{titsiasVariationalLearningInducing2009} which is supported by theoretical guarantees~\cite{burtConvergenceSparseVariational2020,niemanAdaptiveSparseVariational2025}.

The Variational Sparse Gaussian Process chooses the subset of points to minimise the Kullback-Leibler divergence between the approximate and actual posterior. 

Another approach for dealing with the computational bottleneck caused by the high-dimensional kernel (covariance) matrix is based on the Vecchia \cite{vecchiaEstimationModelIdentification1988} approach recently discussed by Katzfuss and Guinness~\cite{katzfussGeneralFrameworkVecchia2021}. The differences to the \BCM are discussed in Section~\ref{ss:BCM}.  This results in a good approximation with substantial sparsity.  However, the approach does not readily lend itself to multiple lengthscale parameters and to parallel implementation on High Performance Computing (HPC) architectures. 
Again, other approaches use decompositions or approximations for the covariance matrix of the \GP~\cite{williamsUsingNystromMethod2000}. These approaches are not suitable for the inference context, since in our application the covariance matrix changes. 

A comprehensive overview of approximation approaches for Gaussian processes with large data sets is provided by Liu et al.~\cite{liuWhenGaussianProcess2020}.

To accommodate large data sets, machine learning is an obvious contender, but requires embedding within an overarching statistical formulation to produce prediction uncertainties.

To avoid being constrained to a limited number of real data points in a high-dimensional context and still get the benefits of the Hierarchical Bayesian Calibration without an excessive penalty in the runtime, we propose --- for the first time --- to combine the Hierarchical Bayesian Calibration with the Bayesian Committee Machine. The Bayesian Committee Machine (\BCM)~\cite{trespBayesianCommitteeMachine2000} splits the \GP of a large data set into multiple small data sets and combines their results by multiplying their likelihood estimates.

Specifically, the BCM is used to speed up evaluation of the likelihood term resulting from the Hierarchical Bayesian Calibration model, making MCMC computationally feasible.

The unnormalised posterior is sampled using the No-U-Turn Hamiltonian Monte Carlo sampler. In the following section, these components and their connections to one another are explained.
 Section~\ref{ss:modelsum} presents the unified calibration formulation with all the components. The hierarchical \BC with \BCM framework is validated for accuracy and scaling in Section~\ref{s:results} against multiple benchmark functions (Section~\ref{s:testproblems}), and the real-world particle accelerator application presented in Section~\ref{s:awa}.

\section{Method}

\subsection{Common Bayesian Calibration}
\label{ss:bayescal}

\BC as introduced by Kennedy and O'Hagan (KOH)~\cite{kennedyBayesianCalibrationComputer2001} infers a model parameter $\theta$ from two data sets. The experimental data set $D_{exp} = (X_{exp},\bm{z}_{exp}) = \{(x_j,z_j)\}_{N_{exp}}$ contains measurements $z_j$ from real-world experiments with corresponding inputs $x_j$. The simulation data set $D_{sim} = (X_{sim},\bm{\theta}_{sim},\bm{y}_{sim}) = \{(x_i,\theta_i,y_i)\}_{N_{sim}}$ contains the outputs $y_i$ produced by the simulator for the inputs $x_i$ and the parameter values $\theta_i$.
Although the inputs $x_{i/j}$ are known for both the experiment and the simulator, the parameter $\theta$ is not known for the experiment beyond a prior $\pi(\theta)$. $N_{exp}$ and $N_{sim}$ are the number of experiments and the number of simulations available. The simulation output $y_j$ and the measurements $z_i$ are modelled as

\begin{align} 
  \nonumber
  y_i &= \eta(x_i, \theta_i)  + \epsilon_i, i=1,\ldots,N_{sim} \\
  z_j &=  \eta(x_j, \theta) + \delta(x_j) + \epsilon_j, j=1,\dots,N_{exp} \nonumber.
\end{align}

The computer code is represented by $\eta(x,\theta)$, the code inadequacies by $\delta(x)$, and the measurement errors by $\epsilon_i$. Surrogates for $\eta(x_i,\theta)$ and $\delta(x_i)$ are required, which need to be inferred in addition to the parameter of interest $\theta$.

 A common approach is to model $\eta(x_i, \theta)$ and $\delta(x_i)$ using \GP and then use MCMC to sample the posterior distribution for the parameter of interest $\theta$.  We follow the concrete description of \BC developed by Higdon et al.~\cite{higdonCombiningFieldData2004} and use the No-U-Turn-Sampler (Section~\ref{ss:nuts}) as the Markov chain sampler.

A \GP~\cite{rasmussenGaussianProcessesMachine2005} models a function $f(x)$ using a mean function $m(x)$ and a covariance function $c(x,x')$, where the mean function can be constructed from a function basis. The latter is often kept very simple by using a constant, i.e.\ zero, which is used here and leads to the simplification in \Eq~\ref{eq:gp}.
The covariance function $c(x,x')$ is used to construct a covariance matrix (\textit{Gram matrix}) $K(X,X)$ of all data points $X$. It measures how close inputs are to the available data points to weigh the impact of the individual data points and calculate the uncertainty with which the prediction can be made. 

When we say a function $f(x), x \in \mathcal{R}^p$ is described as 

\begin{align} 
	f(x) \sim GP(m(x),c(x,x')) \overset{\text{(Zero Mean)}}{=} GP(0,c(x,x')), \label{eq:gp} 
\end{align}

we say that $\begin{bmatrix} \bm{y} \end{bmatrix}$,  the restriction of $f()$ to the $n$ locations encoded in ${n \times p}$ matrix $X$, follows the multivariate normal distribution

\begin{align*}
	\begin{bmatrix} \bm{y} \end{bmatrix} &\sim \mathcal{N}_n\left( \bm{m}(X), K(X,X) + \sigma_\epsilon I\right), & X = \begin{bmatrix}
	    x_1 \\ \vdots \\ x_n
	\end{bmatrix},\label{eq:GPasMVN}
\end{align*}

where $ K(X,X)$ is calculated from all its inputs $X$ using $c(x,x')$ and $\sigma_\epsilon$ is noise in the data. This leads to the conditional mean and variance calculations in Appendix~\ref{ap:pred}.

The covariance function $c(x,x')$ can be chosen according to the prior expectation on the smoothness of the modelled function $f(x)$. A well-tested and common choice 
of covariance function for the use of \GP as a surrogate for reasonably regular functions is the Mat\'ern-$\sfrac{3}{2}$ kernel 
 with the hyperparameters lengthscale $l$ and a rescaling parameter $\lambda$. 

 In the case of multidimensional inputs, the lengthscale hyperparameter can be used to make use of Automatic Relevance Determination~\cite{williamsGaussianProcessesRegression1995}. Here, for inputs $\bm{x} \in \mathcal{R}^p$ different lengthscales $l_i, i=1,\dots,n$ are used for each dimension, leading to the covariance function

\begin{equation}
	k_{\lambda,\bm{l}}(\bm{x},\bm{x'}) = \lambda \left( 1 + \sqrt{ 3\sum_i^p l_i^2 (x_i -x'_i)^2} \right)\exp\left( - \sqrt{3 \sum_i^p l_i^2 (x_i -x'_i)^2 }\right),   
\end{equation}
as shown in more detail in the Appendix~\ref{ap:adr}.

Using this construction, the \Eq~\ref{eq:GPasMVN} gives us the log-likelihood (\Eq~\ref{eq:logpdf}) that the inputs in $X$ and the outputs in $\bm{y}$ are from the same function $f(x)$, and allows us to infer the experimental parameter $\theta$.

Higdon et al.~\cite{higdonCombiningFieldData2004} developed an approach that uses zero mean \GP and combines the whole calibration -- including hyperparameters controlling the covariance models -- into a single posterior distribution that is sampled via Markov chain Monte Carlo.
Ignoring the simulator error for now, the KOH model is just the simulation and experimental data modelled by a \GP and an error term:

\begin{align}
    \nonumber
	z_i &= \eta_{\lambda_{\eta},l_{\eta}} (x_i, \theta) + \epsilon, \label{eq:SimpleKOHGP} \\
\nonumber
\eta_{\lambda_{\eta},l_{\eta}}(x,\theta) &\sim GP\left(\bm{0}, K_{\lambda_{\eta},l_{\eta}}\left(\begin{bmatrix}X_{sim} & \bm{\theta}_{sim} \\ X_{exp} & \bm{\theta^*} \end{bmatrix},\begin{bmatrix}X_{sim} & \bm{\theta}_{sim} \\ X_{exp} & \bm{\theta^*} \end{bmatrix}\right)  \right) \text{, where } \bm{\theta}^* = \begin{bmatrix}
    \vdots \\ \theta^* \\ \vdots
\end{bmatrix}.
\nonumber
\end{align}

The \GP $ \eta_{\lambda_{\eta},l_{\eta}}(x,\theta)$ models the simulator where the parameters $\lambda_{\eta}$ and $l_{\eta}$ are the hyperparameters of the covariance function $c_{\lambda,l}((\cdot,\cdot),(\cdot,\cdot))$.
Given the inputs $X_{sim}$, the parameters $\bm{\theta}_{sim}$, and the outputs $\bm{z}_{sim}$ of the simulator data on the one hand and the inputs $X_{exp}$ and outputs $\bm{y}_{exp}$ on the other hand, it gives us the likelihood of a trial parameter $\theta^*$.

Returning to the full KOH model, including the second \GP  
\begin{align}
	\delta_{\lambda_{\delta},l_{\delta}}(x) &\sim  \GP\left(\bm{0}, K_{\lambda_{\delta},l_{\delta}}\left(X_{exp},X_{exp}\right) \right) \nonumber
\end{align}
to model the simulator inadequacies leads to a simple additional contribution to the covariance matrix. Since the combination of two \GP is again a \GP and in the case of zero mean \GP the covariance matrices just need to be added, hence, 

\begin{align}
	\label{eq:kohGPs}
	\eta_{\lambda_{\eta},l_{\eta}}(x,\theta) + \delta_{\lambda_{\delta},l_{\delta}}(x) + \epsilon  &\sim \GP(\bm{0}, K_z(\theta^*,\phi )) \eqcolon \GP_z, \\
	\text{where } K_z(\theta^*,\phi) &= K_{\eta,\lambda_{\eta},l_{\eta}}(\theta^*) + \begin{bmatrix}
	    0 & 0 \\ 0 & K_{\delta,\lambda_{\delta},l_{\delta}}
	\end{bmatrix} + \sigma_{\epsilon}I \label{eq:addingsigmas} \nonumber \\
	\text{and } \phi &= (\lambda_{\eta},l_{\eta},\lambda_{\delta},l_{\delta},\sigma_{\epsilon}), \nonumber
\end{align}
contains the full calibration model.

The variable $\phi$ is introduced as a collection of hyperparameters in order to simplify the notation, but keeping the dependence of terms on the hyperparameters explicit. These are the hyperparameters $\lambda_{\eta},l_{\eta}$ and $\lambda_{\delta},l_{\delta}$ of the kernel functions used, and the error variance $\sigma_\epsilon$. 

Using $\GP_z$ one obtains the likelihood
\begin{align}
	\label{eq:likelihoodzTheta}
	p(\tilde{\bm{z}} | \theta^*, \phi, X_{sim}, \theta_{sim}, X_{exp}) &= \\
	p(\tilde{\bm{z}} | \theta^*, \lambda_{\eta}, l_{\eta}, \lambda_{\delta}, l_{\delta}, \sigma_{\epsilon}, X_{sim}, \theta_{sim}, X_{exp} ) &= (2\pi)^{-(N_{exp}+N_{sim})/2} |K_z(\theta*,\phi)|^{-\sfrac{1}{2}} \exp \left(- \frac{1}{2} \tilde{\bm{z}}^T (K_z(\theta^*,\phi))^{-1}\tilde{\bm{z}} \right) \nonumber 
\end{align}
that a trial parameter $\theta^*$ and all the hyperparameters in $\phi$ explain the outputs $\bm{z}_{exp}$ and $\bm{y}_{sim}$ in the vector $\tilde{\bm{z}} = \begin{bmatrix}
    y_{sim}& z_{exp}
\end{bmatrix}^T$. 

In order to obtain the posterior probability for the parameters, the likelihood~\Eq ~ \ref{eq:likelihoodzTheta} and the priors are used in Bayes' rule, resulting in an unnormalised posterior for $\theta^*$ and $\phi$

\begin{align}
	\label{eq:posteriorthetaphi}
	p(\theta^*, \phi | \tilde{z}, X_{sim}, \theta_{sim}, X_{exp}) &\propto p(\tilde{\bm{z}} | \theta^*, \phi, X_{sim}, \theta_{sim}, X_{exp}) \pi(\theta^*) \pi(\phi).
\end{align}

Although we have an expression for the likelihood in \Eq~\ref{eq:likelihoodzTheta}, the priors $\pi(\theta)$ and $\pi(\phi)$ need to be chosen.

The prior $\pi(\theta)$ of the parameter of interest depends on the context and prior knowledge of the physical system and can be the uniform distribution $\mathcal{U}(a,b)$ over the range of possible parameters $\theta$. 

For priors of the hyperparameters, we follow the choice of Higdon et al.~\cite{higdonCombiningFieldData2004}. We assume that the components of $\phi$ are independent of each other and, therefore, the prior $\pi(\phi)$ is the product of the individual hyperparameters

\begin{align*}
	\pi(\phi) &= \pi(\lambda_{\eta}) \pi(l_{\eta}) \pi(\lambda_{\delta}) \pi(l_{\delta}) \pi(\sigma_{\epsilon}). 
\end{align*}

For the hyperparameters $\lambda_{\eta}$ and $l_{\eta}$ in the kernel function of $\eta_{\lambda_{\eta},l_{\eta}}(x,\theta)$, the priors are chosen as Gamma distributions. We use gamma priors for the lengthscales as shown below.  The probability density function of a variable following the Gamma distribution is given by 
\begin{align*}
	\pi_{\Gamma(\alpha,\beta)}(x) &= \frac{1}{\Gamma(\alpha) \beta^\alpha} x^{\alpha - 1} e^{-\frac{ x}{\beta}}. \label{eq:pdfGamma} 
\end{align*}

The parameters for the prior of $\lambda_{\eta} \sim \Gamma(5,0.2)$ is chosen such that the expectation value $\mathbb{\lambda_{\eta}} = 1$. The parameters in the priors  
\begin{align*}    
l_{\eta} \sim \Gamma( 7, 3 ) \\
l_{\eta,\theta} \sim \Gamma( 7, 3 )
\end{align*} 
support the data driven automatic relevance determination.

The priors for the hyperparameters 
\begin{align*}
	\lambda_{\delta} &\sim \Gamma(2,500) \\
	l_{\delta,x} &\sim \Gamma( 0.5,2 ) 
\end{align*}
in the covariance function of $\delta_{\lambda_{\delta},l_{\delta}}(x)$ are also chosen as Gamma distributions, but with parameters promoting a flatter \GP closer to zero. In this way, the sampling process favours a good fit of $\eta_{\lambda_{\eta},l_{\eta}}$ rather than absorbing discrepancies into the simulator error $\delta_{\lambda_{\delta},l_{\delta}}$. 

For the error, a value close to zero is expected, which is reflected in the prior distribution $\sigma_{\epsilon} \sim \Gamma(1.0,0.01)$.

Now we can use \Eq~\ref{eq:posteriorthetaphi} to directly sample the parameter $\theta$ at the same time as the hyperparameters in $\phi$ using MCMC. The $\theta$ values in the chain provide realisations from the posterior distribution of one common $\theta$ for all experiments. The variance therefore does not reflect the variance of the parameter between the experiments.

\subsection{Hierarchical Bayesian Calibration}
\label{ss:hierachical}

A hierarchical calibration~\cite{nagelUnifiedFrameworkMultilevel2016} extends standard Bayesian calibration by estimating not a single parameter set, but a parameter ensemble associated with individual experiments. In this setting, latent variables are introduced to characterise the underlying distribution of these experiment-specific parameters, thereby enabling joint inference across multiple experimental contexts.

In order to obtain the hierarchical case from the point estimation, we start to calibrate separate $\theta^*_j$ for each individual measurement $j \in N_{exp}$. Therefore, we are calibrating the whole ensemble 
\begin{align*}
    \bm{\theta} = [\theta_1 \dots \theta_N] \text{ , where } N = N_{exp}. 
\end{align*}
Without loss of generality we assume the original parameters $\theta_i$ to be scalars, but the multidimensional parameter case follows likewise.

Additionally, hierarchical parameters are introduced that are inferred together with the ensemble of $\theta_i$. These hierarchical parameters are latent variables that parametrise the joint hierarchical distribution of the $\theta_i$. 

Let's assume the joint distribution is a normal distribution $\mathcal{N}(\mu_\theta,\sigma_\theta)$ 
then the hierarchical parameters are simply 

$$\mu_\theta,\sigma_\theta \text{ , s.t. } \theta_i \sim \mathcal{N}(\mu_\theta,\sigma_\theta). 
$$

Since the latent variables are calibrated with the ensemble of parameters, this leads to a new parameter vector 
\begin{align}
\vartheta = [\theta_1 \dots \theta_{N_{exp}}, \mu_\theta,\sigma_\theta]. \label{eq:vartheta}
\end{align}
While the $\theta_i$ each keep the original prior $\pi(\theta)$, the introduced latent variables need their own suitable priors.

With this new parameter vector the posterior Equation~\ref{eq:posteriorthetaphi} with just a simple parameter $\theta$ changes for the full parameter vector $\vartheta$ to the posterior 

\begin{align}
	p(\vartheta, \phi | D_{sim}, D_{exp}) &\propto \prod_{i=1}^{N_{exp}}p_{GP_z}(z_{exp,i} | x_{exp,i}, \theta_i, D_{sim}, \phi_\delta, \phi_{sim}) \label{eq:hpkohpart1}
    \\
					      &\times p_{GP_z}(\bm{y}_{sim} | X_{sim}, \phi_{sim}) \label{eq:hpkohpart2}
                          \\
					      &\times \prod_{i=1}^{N_{exp}} p_{\mathcal{N}}(\theta_i | \mu_\theta, \sigma_\theta) \label{eq:hpHierachicalpart}
                          \\
					      &\times \pi(\mu_\theta) \times \pi(\sigma_\theta) \times \pi(\phi_{sim})\times \pi(\phi_{\delta}). \label{eq:hppriors}
\end{align}

\Eq~\ref{eq:hpkohpart1} is Bayes' rule for the individual $\theta_i$, where their likelihoods are calculated using the \GP $p_{GP}$. \Eq~\ref{eq:hpkohpart2} is the contribution of the simulation data and \Eq~\ref{eq:hpHierachicalpart} is the hierarchical prior that connects the latent variables to the parameters $\theta_i$. Finally, \Eq~\ref{eq:hppriors} follows with the priors for the latent variables and the hyperparameters. 

The terms in equation lines~\ref{eq:hpkohpart1} and~\ref{eq:hpkohpart2} are the likelihood terms corresponding to the term $p(\tilde{z} | \theta^*, \phi, D_{sim}, X_{exp})$ in the common \BC. The covariance matrix of the combined \GP is now 
\begin{align*}
    K_h(\bm{\theta},\phi) &= K_{\lambda_{\eta},l_{\eta}}\left(\begin{bmatrix}X_{sim} & \bm{\theta}_{sim} \\ X_{exp} & \bm{\theta} \end{bmatrix},\begin{bmatrix}X_{sim} & \bm{\theta}_{sim} \\ X_{exp} & \bm{\theta} \end{bmatrix}\right)  + \begin{bmatrix}
        0 & 0 \\ 0 & K_{\lambda_\delta, l_\delta}(\begin{bmatrix}
            X_{exp} & \bm{\theta}
        \end{bmatrix},
        \begin{bmatrix}
            X_{exp} & \bm{\theta}
        \end{bmatrix})
     \end{bmatrix} 
     + \sigma^2_\epsilon I 
     \\
     \text{ where } \bm{\theta} &= \begin{bmatrix}
         \theta_1 \\ \vdots \\ \theta_{N_{exp}}
     \end{bmatrix}
\end{align*}
so that the probability density is

\begin{align}
    p_{GP_h}&(\tilde{z} | \bm{\theta}, \phi, X_{exp}, X_{sim},\bm{\theta}_{sim}) 
    = (2\pi)^{ -(N_{exp}+N_{sim})/2} |K_h(\theta,\phi)|^{-\sfrac{1}{2}} \exp \left(- \frac{1}{2} \tilde{\bm{z}}^T (K_h(\bm{\theta},\phi))^{-1}\tilde{\bm{z}} \right). \label{eq:hier_likly}
\end{align}

The parameters of the hierarchical distribution need their own priors. We choose $a$ and $b$ from the uniform prior $\pi(\theta)$ in the ordinary \BC as lower and upper bounds for $\theta_i$ and construct a very flat prior hierarchical distribution $\theta_i \sim \mathcal{N}(\mu_\theta,\sigma_\theta)$. Gelman~\cite{gelmanPriorDistributionsVariance2006} showed that non-uniform priors for hierarchical variances can lead to an underestimation of the inferred variance. Together with the bounds $a$ and $b$ this leads to the priors

\begin{align}
    \mu_\theta &\sim \mathcal{U}(a,b), 
    \label{eq:muthetahier} 
    &\text{and }  
    \sigma_\theta &\sim \mathcal{U}(0,\frac{b-a}{\sqrt{12}} ).
\end{align}

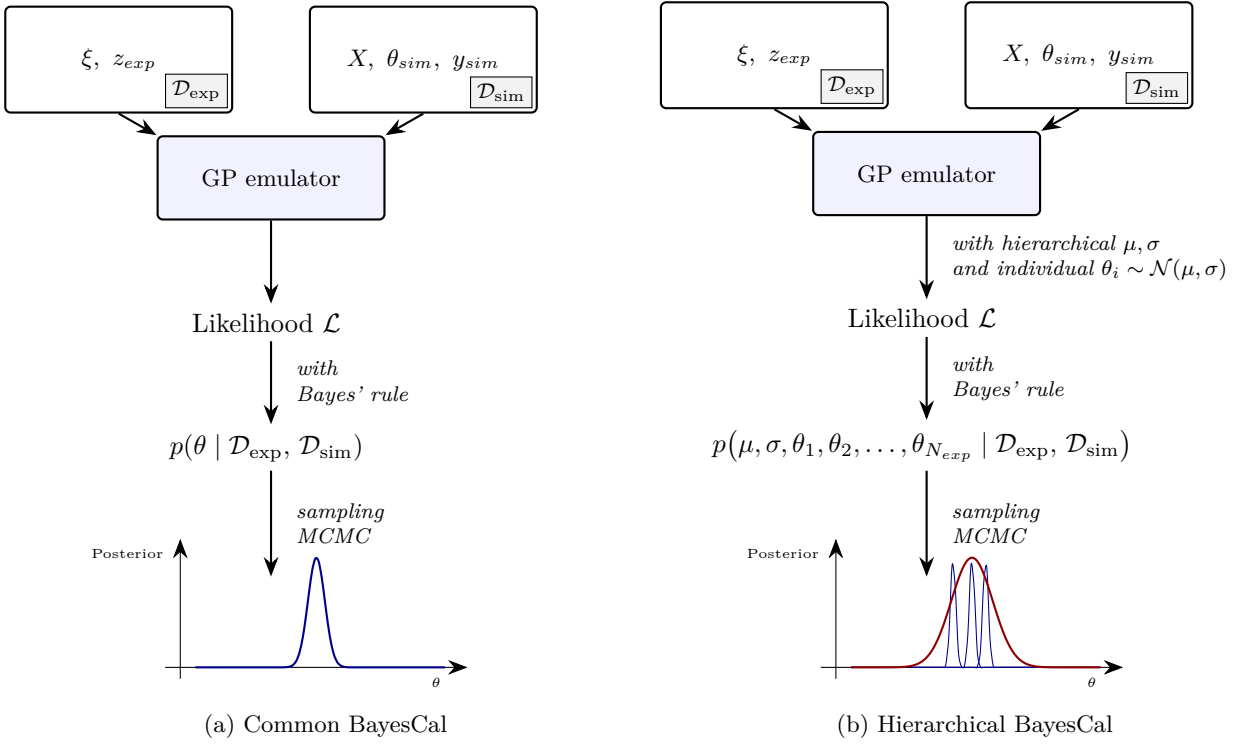
\begin{figure}
    \centering
    \begin{subfigure}[t]{0.50\textwidth}
    \begin{tikzpicture}[
    >={Stealth[length=2.6mm]},
    font=\small,
    data/.style   = {draw, thick, rounded corners=2pt,
                     minimum width=30mm, minimum height=14mm, align=center},
    tag/.style    = {draw, fill=black!5, inner sep=2.5pt, font=\footnotesize},
    proc/.style   = {draw, thick, rounded corners=2pt, fill=blue!5,
                     minimum width=30mm, minimum height=11mm, align=center},
    quant/.style  = {align=center, font=\normalsize},
    arr/.style    = {->, thick},
    note/.style   = {font=\footnotesize\itshape, align=left}
]

\node[data] (exp) {$\xi,\ z_{exp}$};                       
\node[data, right=10mm of exp] (sim) {$X,\ \theta_{sim},\ y_{sim}$}; 

\node[tag, anchor=south east, xshift=-2pt, yshift=2pt]
     at (exp.south east) {$\mathcal{D}_{\mathrm{exp}}$};
\node[tag, anchor=south east, xshift=-2pt, yshift=2pt]
     at (sim.south east) {$\mathcal{D}_{\mathrm{sim}}$};

\coordinate (mid) at ($(exp)!0.5!(sim)$);
\node[proc, below=10mm of mid] (gp) {GP emulator};

\draw[arr] (exp.south) -- (gp.north west);
\draw[arr] (sim.south) -- (gp.north east);

\node[quant, below=11mm of gp] (lik) {Likelihood $\mathcal{L}$ };
\draw[arr] (gp.south) -- (lik.north);

\node[quant, below=11mm of lik] (post)
     {$p\!\left(\theta \mid \mathcal{D}_{\mathrm{exp}},\, \mathcal{D}_{\mathrm{sim}}\right)$ };
\draw[arr] (lik.south) -- node[note, right=2mm] {with\\Bayes' rule} (post.north);

\coordinate (plotc) at ($(post.south) + (0,-26mm)$);
\draw[arr] (post.south) -- node[note, right=2mm] {sampling\\MCMC}
           ($(plotc) + (0,12mm)$);

\begin{scope}[shift={(plotc)}]
  \draw[->] (-1.4,0) -- (2.6,0);                 
  \draw[->] (-1.2,-0.15) -- (-1.2,1.7);          
  \draw[thick, blue!55!black, smooth, samples=80, domain=-1:2.3]
       plot (\x,{1.45*exp(-((\x-0.6)^2)/0.025)}); 
  \node[anchor=west] at (-2.5,1.5) {\tiny Posterior};
  \node[anchor=west] at (2,-0.2) {\tiny $\theta$};
\end{scope}

\end{tikzpicture}
    \caption{Common \BC}
    \end{subfigure}
    \hfill
    \begin{subfigure}[t]{0.49\textwidth}
    \begin{tikzpicture}[
    >={Stealth[length=2.6mm]},
    font=\small,
    data/.style   = {draw, thick, rounded corners=2pt,
                     minimum width=30mm, minimum height=14mm, align=center},
    tag/.style    = {draw, fill=black!5, inner sep=2.5pt, font=\footnotesize},
    proc/.style   = {draw, thick, rounded corners=2pt, fill=blue!5,
                     minimum width=30mm, minimum height=11mm, align=center},
    quant/.style  = {align=center, font=\normalsize},
    arr/.style    = {->, thick},
    note/.style   = {font=\footnotesize\itshape, align=left}
]

\node[data] (exp) {$\xi,\ z_{exp}$};                       
\node[data, right=10mm of exp] (sim) {$X,\ \theta_{sim},\ y_{sim}$}; 

\node[tag, anchor=south east, xshift=-2pt, yshift=2pt]
     at (exp.south east) {$\mathcal{D}_{\mathrm{exp}}$};
\node[tag, anchor=south east, xshift=-2pt, yshift=2pt]
     at (sim.south east) {$\mathcal{D}_{\mathrm{sim}}$};

\coordinate (mid) at ($(exp)!0.5!(sim)$);
\node[proc, below=10mm of mid] (gp) {GP emulator};

\draw[arr] (exp.south) -- (gp.north west);
\draw[arr] (sim.south) -- (gp.north east);

\node[quant, below=11mm of gp] (lik) {Likelihood $\mathcal{L}$ };
\draw[arr] (gp.south) -- node[note, right=2mm] {with hierarchical $\mu,\sigma$\\and individual $\theta_i \sim \mathcal{N}(\mu,\sigma)$} (lik.north);

\node[quant, below=11mm of lik] (post)
     {$p\!\left(\mu,\sigma,\theta_1, \theta_2, \dots, \theta_{N_{exp}} \mid \mathcal{D}_{\mathrm{exp}},\, \mathcal{D}_{\mathrm{sim}}\right)$ };
\draw[arr] (lik.south) -- node[note, right=2mm] {with\\Bayes' rule} (post.north);

\coordinate (plotc) at ($(post.south) + (0,-26mm)$);
\draw[arr] (post.south) -- node[note, right=2mm] {sampling\\MCMC}
           ($(plotc) + (0,12mm)$);

\begin{scope}[shift={(plotc)}]
  \draw[->] (-1.4,0) -- (2.6,0);                 
  \draw[->] (-1.2,-0.15) -- (-1.2,1.7);          
  \draw[thin, blue!55!black, smooth, samples=80, domain=-1:2.3]
       plot (\x,{1.45*exp(-((\x-0.6)^2)/0.0025)}); 
    \draw[thin, blue!55!black, smooth, samples=80, domain=-1:2.3]
       plot (\x,{1.45*exp(-((\x-0.35)^2)/0.0025)}); 
    \draw[thin, blue!55!black, smooth, samples=80, domain=-1:2.3]
       plot (\x,{1.45*exp(-((\x-0.78)^2)/0.0025)}); 
    \draw[thick, red!55!black, smooth, samples=80, domain=-1:2.3]
       plot (\x,{1.45*exp(-((\x-0.6)^2)/0.15)}); 
  \node[anchor=west] at (-2.5,1.5) {\tiny Posterior};
  \node[anchor=west] at (2,-0.2) {\tiny $\theta$};
\end{scope}

\end{tikzpicture}
    \caption{Hierarchical \BC}
    \end{subfigure}
    \caption{Comparing Common (a) and Hierarchical Bayesian Calibration (b): Both combine the experimental and simulation data using a Gaussian Process. This Gaussian process is used to find an expression for the likelihood $\mathcal{L}$ in the unnormalised posterior. While the Common Bayesian Calibration on the left hand side infers just one parameter $\theta$ for all experiments, the Hierarchical \BC on the right hand side infers individual $\theta_i$ for each experiment and a hierarchical distribution $\mathcal{N}(\mu,\sigma)$ (in red), accounting for the variance between experiments.}
    \label{fig:bcvshbc}
\end{figure}

 Unlike running $N_{exp}$ individual calibrations, 
the resulting MCMC chain still uses a single hyperparameter vector $\phi$ across each of the experiments.  
In addition, hierarchical parameters act as a regularising factor, pulling $\theta_i$ towards a joint distribution. At the same time, the hierarchical distribution and its parametrisation introduce a bias. The differences between common and hierarchical \BC are visualised in Figure~\ref{fig:bcvshbc}.

In the case of multidimensional parameters $\theta$ the covariance between the parameter dimensions should be sampled uniformly. These covariances are additional parameters of the hierarchical distribution for multidimensional $\theta$. We use the vine method by Lewandowski et al.~\cite{lewandowskiGeneratingRandomCorrelation2009} to generate a uniformly sampled correlation matrix $C(\theta, \theta')$ by calculating partial conditional correlation terms. Given this valid correlation matrix, the covariance matrix $\Sigma$ is 
\begin{align*}
    \Sigma = Diag(\sigma_\theta) C(\theta, \theta') Diag(\sigma_\theta),
\end{align*}
where $Diag(\sigma_\theta)$ is the diagonal matrix of a vector of $\sigma_\theta$ distributed as in \Eq~\ref{eq:muthetahier}.
Individual $\theta_i \in \mathcal{R}^q$ are distributed as 
\begin{align*}
    \theta_{i,1\dots q} \sim \mathcal{N}\left(\begin{bmatrix}
        \mu_{\theta,1} \\ \vdots \\ \mu_{\theta,q}
    \end{bmatrix}, \Sigma \right)
\end{align*}
where $\mu_{\theta,k}, k \in [1,\dots,q]$ are individual means for each dimension, distributed as in \Eq~\ref{eq:muthetahier}.

\subsection{No-U-Turn Sampler}
\label{ss:nuts}

The No-U-Turn Sampler (NUTS)~\cite{hoffmanNoUTurnSamplerAdaptively2014} is an improvement on the Hamiltonian Monte Carlo (HMC) sampler with the aim to eliminate problem-specific hyperparameters of the sampler itself. 

The HMC sampler introduces an auxiliary momentum $r$ drawn from the normal distribution and treats the logarithmic probability distribution $\log( p_{post}(\theta))$ as pseudo-potential. The sample is now treated like a particle in this Hamiltonian system and evolved in time.

To obtain a new sample the auxiliary Hamiltonian system is evolved using $L$ Leapfrog steps with a step size $\Delta t$. The acceptance 
\begin{align*} 
	\alpha &= \min \left\{ 1, \frac{\exp(\log(p_{post}(\theta^{m}))- \frac{1}{2} \langle r^{m},r^{m} \rangle )} {\exp(\log(p_{post}(\theta^{m-1}))- \frac{1}{2} \langle r^{m-1},r^{m-1} \rangle) } \right\} 
    \end{align*}
of a new sample $\theta^{m}$ is based on the energy change in the auxiliary Hamiltonian system. 
Since Leapfrog integrator is symplectic the pseudo energy of the system is conserved, which ensures a very high probability of accepting new candidates. Further, the time reversibility satisfies the detailed balance condition.

Although the performance of the HMC is strongly influenced by the number of Leapfrog steps taken in each step, the No-U-Turn sampler avoids this by collecting multiple samples from Leapfrog steps forward and backward in time until the distance between the last and first samples starts to decrease. From this list of candidates, the new samples are chosen so that the detailed balance is still satisfied.

An important advantage of HMC and its adaptive variant NUTS over random-walk–based MC methods is their faster convergence, particularly in high-dimensional parameter spaces. A prerequisite for both algorithms is access to the pointwise gradient of the log-posterior, $\grad_{\theta} \log( p_{post}(\theta))$. 
Using the \texttt{Turing.jl} modelling framework~\cite{geTuringLanguageFlexible2018} together with the \texttt{AdvancedHMC.jl} library~\cite{xuAdvancedHMCjlRobustModular2020}, these gradients can be obtained via automatic differentiation. This allows the implementation of custom statistical models without requiring manual gradient derivation or imposing restrictive structural assumptions.

\subsection{Bayesian Committee Machine}
\label{ss:BCM}

The runtime for the evaluation of a \GP with $n$ data points is dominated by the inversion of the covariance matrix and grows with $\mathcal{O}(n^3)$. Therefore, to improve the runtime of a model based on \GP, the number of data points used must be restricted. One common approach is to use reduced-rank approximations of the Gram Matrix~\cite{smolaSparseGreedyGaussian2000,lawrenceFastSparseGaussian2002}. The Bayesian Committee Machine introduced by Tresp~\cite{trespBayesianCommitteeMachine2000} is an alternative. Deisenroth~\cite{deisenrothDistributedGaussianProcesses2015} includes more recent developments and compares common modifications.

Among the reduced-rank approximations, there is the option to reduce the rank of the Gram Matrix by approximating~\cite{williamsUsingNystromMethod2000} it or projecting~\cite{csatoSparseOnLineGaussian2002} onto a smaller subspace. These approximations or projections depend on the Gram Matrix and, therefore, are less suitable in the inference context in which the Gram Matrix changes.

For subset approaches, a subset of data points is selected that contains the most information, while staying within the limits of a given maximal number of data points. The approaches can differ in the strategy used to choose the data points.
The subset approach however limits the model complexity the GP can accurately represent, particularly in higher dimensions.

The Bayesian Committee Machine~\cite{trespBayesianCommitteeMachine2000} and its generalisation~\cite{liuGeneralizedRobustBayesian2018b} keep the number of data points per Gaussian Process small by splitting the data set $D$ into $N_{MBC}$ subsets $D^{(i)}$. This reduces the number of data points per GP to $\frac{n}{N_{MBC}}$, but still makes use of all the data points

$$D = D^{(1)} \cup D^{(2)} \cup \dots D^{(N_{MBC})}.$$
Each member of the Bayesian Committee (MBC) predicts the likelihood of test points given its data subset $D^{(i)}$. These likelihoods can then be combined using
\begin{align}
    \label{eq:BCMcomb}
    p(y^*| D, x^*, \theta^*) \approx \frac{1}{p(y^* | x^*, \theta^*)^{(N_{MBC}-1)}}\prod_{i}^{N_{MBC}} p_{GP}(y^*| D^{(i)}, x^*, \theta^*)
\end{align}
 to approximate the combined likelihood given the complete data set $D$. The product of probabilities needs to be corrected by the $N_{MBC} -1 $ power of the prior of the individual \GP $p_{GP}(y^* |x^*,\theta^*)$. Without this denominator, the variance is underestimated as discussed in Liu et al.~\cite{liuGeneralizedRobustBayesian2018b}.  The \Eq~\ref{eq:BCMcomb} is derived using the chain rule as shown by Tresp~\cite[Section 2]{trespBayesianCommitteeMachine2000}.

Any $p_{GP}(y^* | D_i, x^*,\theta^*)$ is the probability density given by an ordinary \GP evaluated at $y^*$ given the inputs $x^*$ and $\theta^*$, and conditioned on the data subset $D^{(i)}$. The \BCM gives us a way to combine these probabilities to obtain the posterior given all the data sets $D^{(i)}, i=1,\dots,N_{MBC}$. In this manner \BCM is commonly used as a function emulator, but \Eq~\ref{eq:BCMcomb} also allows us to evaluate the likelihood for parameter inference. 

 In the calibration context, the assumption is $|D_{sim}| \gg |D_{exp}|$, which motivates the split of $D_{sim}$, with \newline $D^{(i)}_{sim} = (\bm{y}_{sim}^{(i)}, X^{(i)}_{sim}, \bm{\theta}_{sim}^{(i)})$.

Randomly splitting the simulation data into two subsets
$$D_{sim} = D_{sim}^{(1)} \cup D_{sim}^{(2)},$$
and using 
 \Eq~\ref{eq:BCMcomb}, the likelihood term 
 $p_{GP_z}(\tilde{z} | \bm{\theta}, \phi, D_{sim}, D_{exp})$ in \Eq~\ref{eq:hpkohpart1}  transforms into

\begin{align}
    p(\tilde{z} | \bm{\theta}, \phi, D_{sim}, D_{exp}) 
    = p(\bm{z}_{exp} ,\bm{y}_{sim}^{(1)},\bm{y}_{sim}^{(2)}
     | \bm{\theta}, \bm{x}_{exp}, X^{(1)}_{sim}, \bm{\theta}_{sim}^{(1)}, X^{(2)}_{sim}, \bm{\theta}_{sim}^{(2)}, \phi) \label{eq:BCMcomb1}\\
    \stackrel{\text{(Eq~\ref{eq:BCMcomb})}}{\approx}
    \frac{p(\bm{z}_{exp},\bm{y}_{sim}^{(1)} | \bm{\theta}, \bm{x}_{exp}, X^{(1)}_{sim}, \bm{\theta}_{sim}^{(1)},\phi) 
    p(\bm{z}_{exp},\bm{y}_{sim}^{(2)} | \bm{\theta}, \bm{x}_{exp}, X^{(2)}_{sim}, \bm{\theta}_{sim}^{(2)},\phi)}{p(\bm{z}_{exp}|\bm{x}_{exp},\bm{\theta},\phi)}. 
    \label{eq:BCMlikelihood}
\end{align}

The equality in \Eq~\ref{eq:BCMcomb1} only holds if the data set on which the posterior is conditioned is conditionally independent. This is true for $D_{sim}^{(i)}$ as long as the subsets are drawn from $D_{sim}$ independently. Strictly speaking, this is not true for $D_{exp}$, which is not conditionally independent from itself.

The denominator in \Eq~\ref{eq:BCMlikelihood}, correcting for variance underestimation, is

\begin{align*}
    p(\bm{z}_{exp}|\bm{x}_{exp},\bm{\theta},\phi) &= (2\pi)^{-(N_{exp})/2} |W|^{-\sfrac{1}{2}} \exp \left( -\frac{1}{2}\tilde{\bm{z}}_{exp}^T W^{-1}\tilde{\bm{z}}_{exp} \right) \\
    W &= K_{\lambda_{\eta},l_{\eta}}\left(\begin{bmatrix}X_{exp} & \bm{\theta} \end{bmatrix},\begin{bmatrix}X_{exp} & \bm{\theta} \end{bmatrix} \right).
\end{align*}

\begin{figure}
    \centering
    
    \begin{tikzpicture}[
    >={Stealth[length=2.6mm]},
    font=\small,
    data/.style   = {draw, thick, rounded corners=2pt, fill=white,
                     minimum width=30mm, minimum height=10mm, align=left},
    tag/.style    = {draw, fill=black!5, inner sep=2.5pt, font=\footnotesize},
    proc/.style   = {draw, thick, rounded corners=2pt, fill=blue!5,
                     minimum width=32mm, minimum height=12mm, align=center},
    quant/.style  = {align=center, font=\normalsize},
    arr/.style    = {->, thick},
    lin/.style    = {-, thick},
    note/.style   = {font=\footnotesize\itshape, align=left}
]

\node[data] (exp) {$\xi,\ z$};                       
\node[data, right=10mm of exp] (sim) {$X,\ \theta,\ y$}; 

\node[tag, anchor=south east, xshift=-2pt, yshift=2pt]
     at (exp.south east) {$\mathcal{D}_{\mathrm{exp}}$};
\node[tag, anchor=south east, xshift=-2pt, yshift=2pt]
     at (sim.south east) {$\mathcal{D}_{\mathrm{sim}}$};

\node[data, below=9mm of sim] (subs) {$X,\ \theta,\ y$};
\node[tag, anchor=south east, xshift=-2pt, yshift=2pt]
     at (subs.south east) {$\mathcal{D}_{\mathrm{sim}}^{(i)}$};
\stackbehind{subs}{white}
\draw[arr] (sim.south) -- node[note, right=5mm] {random split} (subs.north);
\draw[arr] (sim.south) -- ($(subs.north)+(0.2,0.2)$);
\draw[arr] (sim.south) -- ($(subs.north)+(0.4,0.4)$);

\coordinate (mid) at ($(exp)!0.5!(sim)$);
\node[proc, below=30mm of mid] (gp) {GP emulators};
\stackbehind{gp}{blue!5}

\draw[arr] (exp.south)  -- (gp.north west);   
\draw[arr] (subs.south) -- (gp.north east);   
\draw[arr] ($(subs.south)+(0.5,0)$) -- ($(gp.north east)+(0.2,-0.2)$); 
\draw[arr] ($(subs.south)+(1,0)$) -- ($(gp.north east)+(0.4,-0.4)$); 

\node[proc, below=8mm of gp] (lik) {$\mathcal{L}_1,\ \dots,\ \mathcal{L}_n$};
\stackbehind{lik}{blue!5}
\draw[arr] (gp.south) -- (lik.north);
\draw[arr] ($(gp.south)+(0.2,0)$) -- ($(lik.north)+(0.2,0.2)$);
\draw[arr] ($(gp.south)+(0.4,0)$) -- ($(lik.north)+(0.4,0.4)$);

\node[quant, below=15mm of lik] (clik) {Likelihood $\mathcal{L}$};
\draw[arr] (lik.south) -- node[note, right=5mm] {all\\combined\\with BCM} (clik.north);
\draw[arr] ($(lik.south)+(0.2,0)$) --  ($(clik.north) +(0.07,0)$);
\draw[arr] ($(lik.south)+(0.4,0)$) --  ($(clik.north)+(0.14,0)$);

\node[quant, below=5mm of clik] (post)
     {$p\!\left(\theta \mid \mathcal{D}_{\mathrm{exp}},\, \mathcal{D}_{\mathrm{sim}}\right)$};
\draw[arr] (clik.south) --  (post.north);

\coordinate (plotc) at ($(post.south) + (0,-20mm)$);
\draw[arr] (post.south) -- ($(plotc) + (0,12mm)$);

\begin{scope}[shift={(plotc)}]
  \draw[->] (-1.4,0) -- (2.6,0);                 
  \draw[->] (-1.2,-0.15) -- (-1.2,1.7);          
  \draw[thick, blue!55!black, smooth, samples=80, domain=-1:2.3]
       plot (\x,{1.45*exp(-((\x-0.6)^2)/0.18)}); 
  \node[anchor=west] at (-2.5,1.5) {\tiny Posterior};
  \node[anchor=west] at (2,-0.2) {\tiny $\theta$};
\end{scope}

\end{tikzpicture}
    \caption{Bayesian Calibration with Bayesian Committee Machine: After splitting the simulation data $D_{sim}$ these smaller data sets are used to express parameter likelihoods, which are then combined into one likelihood using \Eq~\ref{eq:BCMcomb}. This likelihood is used to express the posterior for $\theta$.}
    \label{fig:schema_BCM}
\end{figure}
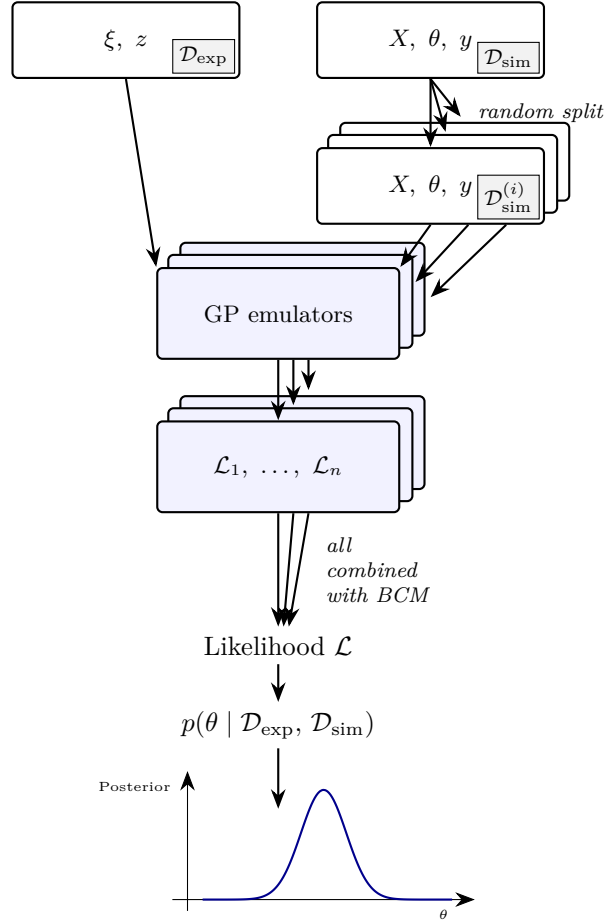

In practice, this allows us to calculate the likelihoods $p(\bm{z}_{exp},\bm{y}_{sim}^{(i)} | \bm{\theta}, \bm{x}_{exp}, X^{(i)}_{sim}, \bm{\theta}_{sim}^{(i)},\phi)$ for the different data sets $D_{sim}^{(i)}$ separately and multiply the individual probabilities. Splitting into more \MBCS follows the same recipe. Although this already has a smaller asymptotic runtime, another beneficial aspect of \BCM is that it lends itself to parallelisation, since each product term in \Eq~\ref{eq:BCMlikelihood} can be calculated separately. Figure~\ref{fig:schema_BCM} shows the flow of data and how the \BCM integrates into \BC.

\subsubsection{Validity of BCM}
\label{ss:valofBCM}

The validity of the \BCM approximation hinges on the conditional independence of the data sets $D_{sim}^{(i)}$, which allows the use of the chain rule (details in Tresp~\cite[Section 2]{trespBayesianCommitteeMachine2000}). In this case the prediction is Bayes optimal if the number of query points (here equivalent to number of experiments) equals the effective degrees of freedom (Tresp~\cite[Section 4]{trespBayesianCommitteeMachine2000}). However, for a lower number of query points, there are no proven convergence guarantees known to the authors, and the degrees of freedom of the emulated function are generally not known. Tresp did show an empirical decrease in the error for an increasing number of query points. For cases in which stronger guarantees are important, an approach with a more thorough theoretical grounding in the literature, such as Variational Sparse Gaussian Processes, might be better suited.

One drawback of the \BCM is that predictions may lack accuracy when compared to the standard setting where the emulator GP conditions on all of the simulation output at once.  In addition, estimated uncertainty may be inaccurate due to non-informative committee members and not accurately accounting for dependence between committee members.  Also, ``outliers'' resulting from some simulations (e.g. unconverged iterations, input errors, etc.) may have an outsized impact on \BCM-based predictions.

In the calibration context for our accelerator application, these shortcomings are less pronounced. The simulation output $D_{sim}$ is from a high accuracy, thoroughly tested computational model, using a common input specification to model the Argonne Wakefield Accelerator. We also see smooth, stable output changes as the accelerator parameters are varied. Thus GP predictions resulting from different committee members will yield accurate predictions of untried model runs.

\subsubsection{Adaptations of \BCM }

Although here the split of $D_{sim}$ into $D_{sim}^{(i)}$ is done randomly, some implementations introduce a deterministic domain decomposition as in the Generalised Robust BCM (GRBCM)~\cite{liuGeneralizedRobustBayesian2018b}. These domain decompositions of the data set produce less noisy local experts but are not conditionally independent. The GRBCM adds random global data sets to maintain conditional independence. Especially in problem settings with obvious decompositions (e.g. phase transitions or different lengthscales over the domain), it could be worth changing the data set split to improve the performance.

While not considered in this paper,
further computational speed-up could be obtained by combining BCM with other approximation approaches
-- such as those introduced in Section~\ref{intro}. 
This is possible since the individual \MBCS are again just GPs.
For example, the Sparse Bayesian Committee Machine~\cite{willowSparseBayesianCommittee2025} shows how \BCM can be combined with Variational Sparse GP for prediction. 
For the applications of this paper, the \MBCS are kept as ordinary, simple GPs.

We find that with this basic incorporation of \BCM into Bayesian Calibration, larger sets of simulation data can be used when available, without a large effect on the sampling runtime.

\subsection{Model Summary}
\label{ss:modelsum}

\begin{figure}
    \centering
    \begin{tikzpicture}[
    >={Stealth[length=2.6mm]},
    font=\small,
    data/.style   = {draw, thick, rounded corners=2pt, fill=white,
                     minimum width=40mm, minimum height=10mm, align=center},
    tag/.style    = {draw, fill=black!5, inner sep=2.5pt, font=\footnotesize},
    proc/.style   = {draw, thick, rounded corners=2pt, fill=blue!5,
                     minimum width=32mm, minimum height=12mm, align=center},
    quant/.style  = {align=center, font=\normalsize},
    arr/.style    = {->, thick},
    lin/.style    = {-, thick},
    note/.style   = {font=\footnotesize\itshape, align=left}
]

\node[data] (exp) {$\xi,\ z_{exp}$};                       
\node[data, right=40mm of exp] (sim) {$X,\ \theta_{sim},y_{sim}$}; 

\node[tag, anchor=south east, xshift=-2pt, yshift=2pt]
     at (exp.south east) {$\mathcal{D}_{\mathrm{exp}}$};
\node[tag, anchor=south east, xshift=-2pt, yshift=2pt]
     at (sim.south east) {$\mathcal{D}_{\mathrm{sim}}$};

\node[data, below=10mm of sim] (subs) {$X^{(k)},\ \theta^{(k)},\ y^{(k)}$};
\node[tag, anchor=south east, xshift=-2pt, yshift=2pt]
     at (subs.south east) {$\mathcal{D}_{\mathrm{sim}}^{(k)}$};
\stackbehind{subs}{white}
\draw[arr] (sim.south) -- node[note, right=5mm] {random split} (subs.north);
\draw[arr] (sim.south) -- ($(subs.north)+(0.2,0.2)$);
\draw[arr] (sim.south) -- ($(subs.north)+(0.4,0.4)$);

\coordinate (mid) at ($(exp)!0.5!(sim)$);
\node[proc, below=30mm of mid] (gp) {Gaussian process\\emulators};
\stackbehind{gp}{blue!5}

\draw[arr] (exp.south)  -- (gp.north west);   
\draw[arr] (subs.south) --  (gp.north east);   
\draw[arr] ($(subs.south)+(0.5,0)$) -- ($(gp.north east)+(0.2,-0.2)$); 
\draw[arr] ($(subs.south)+(1,0)$) -- ($(gp.north east)+(0.4,-0.4)$); 

\node[proc, below=15mm of gp] (lik) {$\mathcal{L}_1,\ \dots,\ \mathcal{L}_n$};
\stackbehind{lik}{blue!5}
\draw[arr] (gp.south) -- node[note, right=5mm] {with hierarchical $\mu,\sigma$\\and individual $\theta_i \sim \mathcal{N}(\mu,\sigma)$} (lik.north);
\draw[arr] ($(gp.south)+(0.2,0)$) -- ($(lik.north)+(0.2,0.2)$);
\draw[arr] ($(gp.south)+(0.4,0)$) -- ($(lik.north)+(0.4,0.4)$);

\node[quant, below=15mm of lik] (clik) {Likelihood $\mathcal{L}$};
\draw[arr] (lik.south) -- node[note, right=5mm] {all\\combined\\with BCM} (clik.north);
\draw[arr] ($(lik.south)+(0.2,0)$) --  ($(clik.north) +(0.07,0)$);
\draw[arr] ($(lik.south)+(0.4,0)$) --  ($(clik.north)+(0.14,0)$);

\node[quant, below=11mm of clik] (post)
     {$p\!\left(\mu,\sigma,\theta_1, \theta_2, \dots, \theta_{N_{exp}} \mid \mathcal{D}_{\mathrm{exp}},\, \mathcal{D}_{\mathrm{sim}}\right)$};
\draw[arr] (clik.south) -- node[note, right=2mm] {with\\Bayes' rule} (post.north);

\coordinate (plotc) at ($(post.south) + (0,-26mm)$);
\draw[arr] (post.south) -- node[note, right=2mm] {sampling\\with NUTS}
           ($(plotc) + (0,12mm)$);

\begin{scope}[shift={(plotc)}]
  \draw[->] (-1.4,0) -- (2.6,0);                 
  \draw[->] (-1.2,-0.15) -- (-1.2,1.7);          
  \draw[thin, blue!55!black, smooth, samples=80, domain=-1:2.3]
       plot (\x,{1.45*exp(-((\x-0.6)^2)/0.0025)}); 
    \draw[thin, blue!55!black, smooth, samples=80, domain=-1:2.3]
       plot (\x,{1.45*exp(-((\x-0.35)^2)/0.0025)}); 
    \draw[thin, blue!55!black, smooth, samples=80, domain=-1:2.3]
       plot (\x,{1.45*exp(-((\x-0.78)^2)/0.0025)}); 
    \draw[thick, red!55!black, smooth, samples=80, domain=-1:2.3]
       plot (\x,{1.45*exp(-((\x-0.6)^2)/0.15)}); 
  \node[anchor=west] at (-2.5,1.5) {\tiny Posterior};
  \node[anchor=west] at (2,-0.2) {\tiny $\theta$};
\end{scope}

\end{tikzpicture}
    \caption{Full Hierarchical Bayesian Calibration with Bayesian Committee machine: 
    First, the simulation data set $D_{sim}$ is split randomly. Next each $D^{(k)}_{sim}$ is combined with the experimental data $D_{exp}$ to express the likelihood $\mathcal{L}_k$ for all $\theta_i$ as well as the hierarchical parameters $\mu$ and $\sigma$ given $D_{exp}$ and $D^{(k)}_{sim}$. These likelihoods $\mathcal{L}_k$ are combined into one likelihood using Equation~\ref{eq:BCMcomb}. Finally, the unnormalised posterior is sampled using the No-U-Turn Sampler.}
    \label{fig:schema_full}
\end{figure}
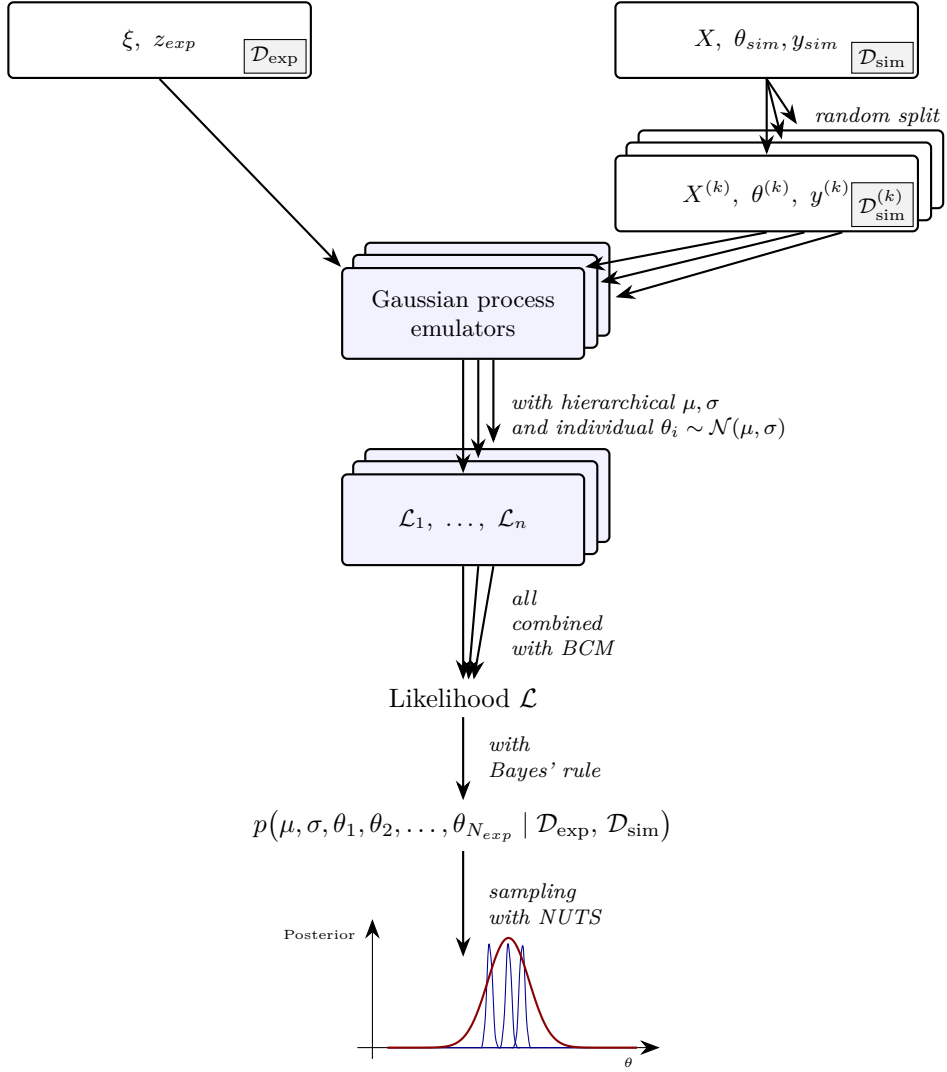

Figure~\ref{fig:schema_full} shows all components and how they interlock in the full framework. 

We use the KOH model, but in combining the kernel matrices, we model the right hand side by a single Gaussian Process (\Eq~\ref{eq:kohGPs}):

\begin{align}
    \begin{pmatrix} \bm{z}_{exp} \\ \bm{y}_{sim} \end{pmatrix} \sim GP_z(\bm{0},\Sigma_z(\vartheta,\Phi_{\delta},\sigma_\epsilon) | \Phi_{\eta}, D_{sim}, D_{exp}) \label{eq:finalmodel}.
    \end{align} 

The vector inferred is $[\theta_1, \dots, \theta_{N_{exp}}, \mu_{post}, \sigma_{post}, \Phi_{\delta}, \sigma_\epsilon ]$, which is the extended vector for Hierarchical Calibration $\vartheta$ (\Eq~\ref{eq:vartheta}), the hyperparameters for the simulator error $\Phi_\delta$ and uncertainty $\epsilon$. The data sets are the simulation data $D_{sim}$ and the observation data $D_{exp}$. Hyperparameters $\Phi_{\eta}$ are optimised beforehand by maximising the likelihood on $D_{sim}$ and are no longer a dependency of $\Sigma_z$.

This \GP (\Eq~\ref{eq:finalmodel}) is then randomly split into $N_{MBC}$ members of a \BCM obtaining the full posterior probability (cf. \Eq~\ref{eq:BCMlikelihood} and \Eq~\ref{eq:hpkohpart1}). 

The introduced parameters all need priors. For the hyperparameters we follow the suggestion by Higdon et al.~\cite{higdonCombiningFieldData2004}, for the calibration parameters $\theta_i$ we assume lower limit $a$ and upper limit $b$. This leads to the priors for the hierarchical parameters $\mu_{post}$ and $\sigma_{post}$ as described in Section~\ref{ss:hierachical}. With 
$\theta_i$ the hierarchical prior and $dim(\bm{l}_\delta) = dim(\bm{l}_{sim}) - dim(\theta_i)$ we get

\begin{align}
    \theta_i &\sim \mathcal{N}(\mu_\theta, \sigma_\theta) \\
    \lambda_{sim,k} &\sim \Gamma(5,0.2) \nonumber \\
    \bm{l}_{sim,k} &\sim [ \Gamma(7,3), \Gamma(7,3), \dots ] \nonumber \\
    \lambda_{\delta,k} &\sim \Gamma(2,500) \nonumber \\
    \bm{l}_{\delta,k} &\sim [ \Gamma(0.5,2), \Gamma(0.5,2), \dots ] \nonumber\\
    \mu_\theta &\sim \mathcal{U}(a, b) \\
    \sigma_\theta &\sim \mathcal{U}\left(0,\frac{b-a}{\sqrt{12}}\right) \nonumber
\end{align}
and $k \in [1... N_{MBC}]$ iterates over the members of \BCM with individual hyperparameters. The difference in the dimensions of $\bm{l}_\delta$ and $\bm{l}_{sim}$ follows from the difference in the kernel functions. Details are shown in Appendix~\ref{ap:sigma}.

\section{Results}
\label{s:results}
The proposed approach is first evaluated on two standard benchmark problems — the Park function and the cantilever beam function — and subsequently applied to a real system, namely the Argonne Wakefield Accelerator Beam Test Facility introduced earlier.

\subsection{Test Problems and Application}
\label{s:testproblems}

\subsubsection{Park Function}
\label{ss:parkfunction}

The Park function (\Eq~\ref{eq:park_org}), proposed by Park~\cite{coxStatisticalMethodTuning2001} for calibration tasks, serves as a closed form, multidimensional test function on which to test the implementation and modifications to the ordinary KOH \BC. 

\begin{align}
    f_{\text{Park}}(\bm{x}) &= \frac{2}{3} e^{x_1 + x_2} - x_4 \sin(x_3) + x_3 \label{eq:park_org} \\
    f(\bm{x},\theta) &= \frac{2}{3} e^{\theta + x_1} - x_2 \sin(x_3) + x_3 + \epsilon \cos(100 x_3) \label{eq:park_tfunc} \\
    f_{sim}(\bm{x},\theta) &= \frac{2}{3} e^{\theta + x_1} - x_2 \sin(x_3) + x_3 \label{eq:park_simfunc} 
\end{align}

The first input dimension $x_1$ is chosen as parameter $\theta$ which is inferred in the calibration. In order to test all components of the setup a simulator error $\epsilon \cos(100 x_3)$, which is missing from the simulator function is added. The function~$f(\bm{x},\theta)$~(\Eq~\ref{eq:park_tfunc}) is the modified Park function that describes the real-world process, while the function~$f_{sim}$~(\Eq~\ref{eq:park_simfunc}) mimics a simulator code of $f(\bm{x},\theta)$. 

\begin{table}[ht]
\centering
\caption{Data and distributions used for the Park function. $N$ is the number of data points in the data sets for different experiments, while the inputs $\bm{x}$, parameter $\theta$ and output $z$ columns state how the respective data is generated.}
 \begin{tabular}{| l | c | c | c | c |} 
 \hline
	 & $N = |D_i|$ &  $\bm{x} / \bm{\xi}$ & $\theta$ & $z$ \\
 \hline
     $D_{sim}$ & $N_{sim} \in [ 25, 6400]$ & $x_{1,2,3} \sim \mathcal{U}(0,1)$ & $\theta_{sim} \sim \mathcal{U}(0,1)$ & $z = f_{sim}(x,\theta_{sim})$ \\
	       & (varies) & (uniform for each dimension) & (uniform) & \\
	\hline
     $D_{exp}$ & $N_{exp} = 10$ & $\xi_{1,2,3} \sim \mathcal{U}(0,1)$ & $ \theta_{true} \sim \mathcal{N}(0.65,0.025)$ & $\zeta = f(\bm{\xi},\theta_{true})$\\
	       &  & (uniform for each dimension) & ( \textbf{to infer} ) & \\
     \hline
    \end{tabular}
     
	    \label{tbl:park_data}
\end{table}

The inputs $x, \xi$ and the simulation parameter $\theta_{sim}$ are drawn from Uniform distributions, while the observation parameter $\theta_{true}$ is drawn from the normal distribution $\mathcal{N}(0.65,0.025)$. This normal distribution is the ground truth for the calibration process. Table~\ref{tbl:park_data} details how the data sets $D_{sim}$ and $D_{exp}$ are generated for the calibration of $\theta_{true}$ in the Park function. 

As prior for $\theta$ the range $[0.5,1]$ is used. The assumption is that it is reasonably certain that $\theta$ will be found in this region.

\subsubsection{Cantilever Beam Functions}

The cantilever beam functions~\cite{eldredInvestigationReliabilityMethod2007} --- used to test calibration and uncertainty quantification methods --- are another suitable test case. They describe the displacement (\Eq~\ref{eq:displacement}) and stress (\Eq~\ref{eq:stress}) in a beam based on horizontal $X$ and vertical load $Y$. In contrast to the Park function, it showcases multidimensional calibration parameters and multiple outputs. 

\begin{align}
    D(E,X,Y) = \frac{4 L^3}{Ewt}\sqrt{\left(\frac{Y}{t^2} \right)^2 + \left(\frac{X}{w^2}\right)^2} \label{eq:displacement}\\
    S(X,Y) = \frac{600Y}{wt^2} + \frac{600X}{w^2t} \label{eq:stress}
\end{align}

In this calibration setting, the input vector $\bm{x}$ consists of the applied loads $X,Y$, while the parameter vector 
$\bm{\theta}$ contains the beam width 
$w$ and thickness $t$. All variables are rescaled such that inputs and outputs have a comparable numerical magnitude. Analogously to the Park-function example, a cosine perturbation is added to construct the simulated observations, resulting in the functions
\begin{align}
    f_{D,exp}(\bm{x},\bm{\theta}) &= \frac{4 \times 100^3}{\SI{2.9e7} \times \theta_1  \theta_2} \sqrt{\frac{x_2^2}{4 \theta_2^4} + \frac{x_1^2}{\theta_1^4} } + \epsilon \cos(100 x_1) \label{eq:f_d_exp}\\
    f_{D,sim}(\bm{x},\bm{\theta}) &= \frac{4 \times 100^3}{\SI{2.9e7} \times \theta_1  \theta_2} \sqrt{\frac{x_2^2}{4 \theta_2^4} + \frac{x_1^2}{\theta_1^4} }  \\
    f_{S,exp}(\bm{x},\bm{\theta}) &= \frac{x_2 }{2 \theta_1 \theta_2^2} + \frac{ x_1}{\theta_1^2 \theta_2} + \frac{\epsilon}{2} \cos(50 x_1)\\
    f_{S,sim}(\bm{x},\bm{\theta}) &=  \frac{x_2 }{2 \theta_1 \theta_2^2} + \frac{ x_1}{\theta_1^2 \theta_2} \label{eq:f_s_sim}.
\end{align}

In this adapted version, the inputs $x_{1,2}$ take values in $[0,1]$. The parameters $\theta_{1,2}$ are in $[0.1,1]$ to avoid division by very small values. In addition, for all outputs their logarithmic value $\log(f_{\cdot,\cdot}(x,\theta))$ is used to keep them within a few orders of magnitude. 

\subsubsection{Argonne Wakefield Accelerator Beam Test Facility} 
\label{s:awa}
\begin{figure}[ht!]
    \centering   
    \includegraphics[width=0.75\textwidth]{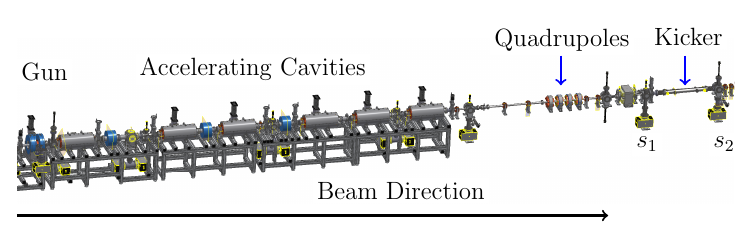}
    \caption{Overview of the Argonne Wakefield Accelerator beam test facility. In this work we consider the very first part of the facility, labelled {\tt Gun}. This is the actual electron source of the facility. }
    \label{fig:awa}
\end{figure}

As a concrete application, the Argonne Wakefield Accelerator~(AWA)~\cite{condeResearchProgramRecent2017} beam test facility is considered. Figure~\ref{fig:awa} shows a schematic overview of the facility. 

We focus here on the electron source subsystem (labelled Gun in Figure \ref{fig:awa}) and calibrate a single parameter: the accelerating field amplitude $V_{gun}$. Although this quantity is physically important, it cannot be adjusted directly and must be inferred through its indirect relationships (with the klystron settings), making it a suitable target for calibration. The example is intentionally low-dimensional and serves as a proof-of-concept for extending the approach to additional components or the full accelerator model. For the present study, synthetic observations generated with the OPAL simulation framework~\cite{adelmannOPALVersatileTool2019} are used instead of experimental measurements. A dedicated measurement campaign is scheduled for 2026, at which point these simulated observations will be replaced by empirical data from the electron source and, potentially, the full facility.

The experiment tracks the electron beam from its emission at the gun to its first diagnostic screen, where measurable beam properties such as density and transverse size are recorded. The gun is operated using two control inputs: the injection amplitude 
$V_{gun}$ and the injection phase $\phi_{gun}$. Downstream, the beam is shaped using three solenoids (focusing, matching, and bucking), each regulated by its respective current 
$I_{foc}$, $I_{match}$, $I_{buck}$.

This configuration reflects a typical initial commissioning procedure at AWA and similar accelerator facilities, where the goal is to identify operating regions that yield a stable, high-quality beam. In practice, this process is repeated sequentially at subsequent monitoring stations along the beamline, with each additional element contributing new tunable parameters and control settings. As the beam propagates downstream, the dimensionality and tuning complexity increase correspondingly.

\subsection{Asymptotic Runtime of Bayesian Calibration}

\begin{figure}[ht]
    \centering
    \includegraphics[width=0.75\textwidth]{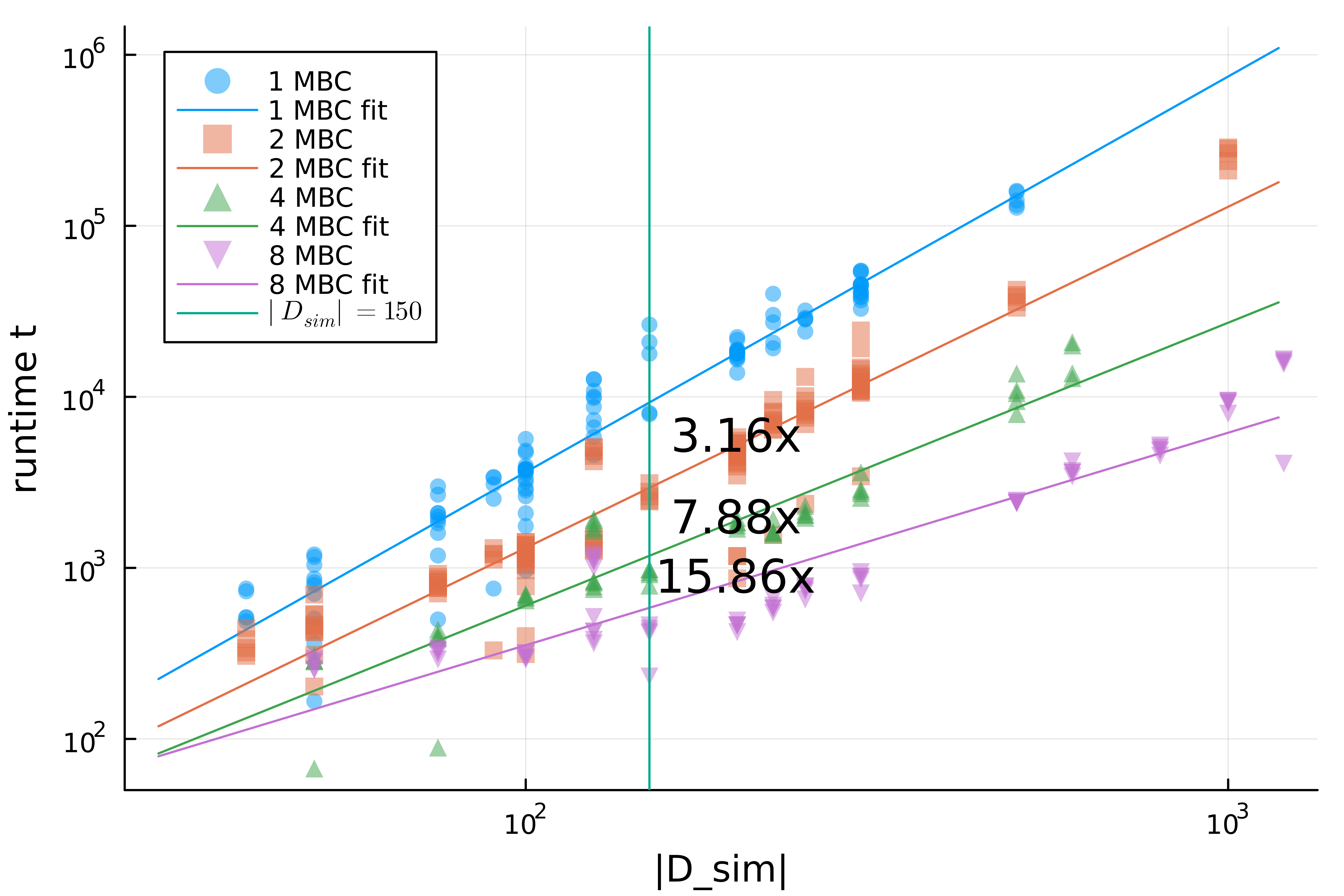}
    \caption{Impact of data set size on the runtime of the \BC. The data series 1 MBC shows the scaling for the base case of the cantilever beam functions with no \BCM, while 2, 4, and 8 MBC show the case for the use of a different number of parallelised \BCM. All calibrations used a chain length of 100. In black the annotations show the speed-up compared to one \MBC at $|D_{sim}| = 150$.}
    \label{fig:kohOdataset}
\end{figure}

The runtime of \BC depends on the dimensionality of the problem and the size of the data set. While the dimensionality of the problem is given, there is a choice in the number of data points used. 
Once $|D_{\rm sim}| \gtrsim 30$, the inversion of the covariance matrix in the \GP dominates the runtime of a Monte Carlo step and, therefore, the total runtime. Figure~\ref{fig:kohOdataset} shows the runtime of the hierarchical \BC for different data set sizes with 1, 2, 4 and 8 \MBCS for the cantilever beam functions.

Figure~\ref{fig:kohOdataset} shows the impact of different data set sizes on the runtime for the use of a different number of parallelised members of the Bayesian Committee. The case without the use of \BCM (1 \MBC) the runtime is $\mathcal{O}(n^{2.31})$. For the use of two \MBCS this falls to $\mathcal{O}(n^{1.99})$ and $\mathcal{O}(n^{1.24})$ for eight \MBCS. In the case of given data set size $|D_{sim}| = 150$ this means a speed-up over no \BCM of 3.16 for two \MBCS and 15.86 for eight \MBCS.

\subsection{Estimate Distribution using Hierarchical Parameters}

In contrast to ordinary \BC, by calibrating individual parameters for each experiment and adding hierarchical parameters as described in Section~\ref{ss:hierachical}, the parameter distribution can be recovered. This is exemplified using the Park function (Section~\ref{ss:parkfunction}) and the specifications in Table~\ref{tbl:parkfunction_recovery_specifications}.

\begin{table}[ht]
\centering
\caption{Specifications used to show the recovery of a given parameter distribution using the hierarchical \BC.}
 \begin{tabular}{|l | c | c | c | c | c| c|} 
 \hline
 & \textbf{$N_{MBC}$} & NUTS-MC chain length & \textbf{$P(\theta_{true})$} & \textbf{$|D_{sim}|$} & \textbf{$|D_{exp}|$} & Prior $\theta$ \\ 
 \hline
     \textbf{Specifications} & 4 & 500 & $\mathcal{N}(0.59895,0.02219)$ & 200 & 10 & $\mathcal{U}(0.5,1)$ \\ 
 \hline
 \end{tabular}
 
	\label{tbl:parkfunction_recovery_specifications}
\end{table}

\begin{figure}[ht]
    \centering
    \includegraphics[width=0.75\textwidth]{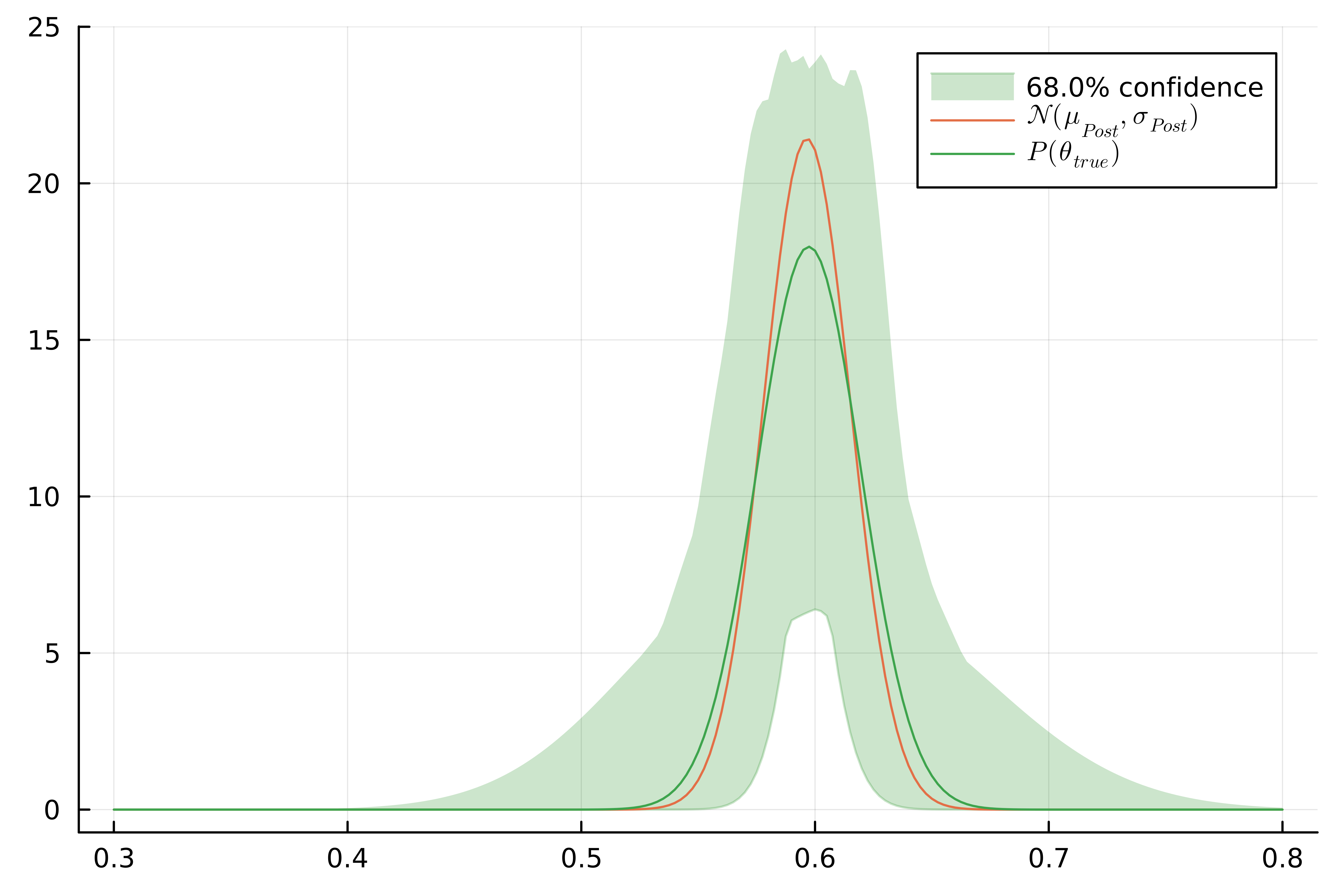}
    \caption{Inferred PDF $\mathcal{N}(\mu_{Post},\sigma_{Post})$ of the parameter $\theta$ including uncertainty compared to true Gaussian PDF $P(\theta_{true})$}
    \label{fig:gaussian_pdf_theta_park}
\end{figure}

Figure~\ref{fig:gaussian_pdf_theta_park} 
 compares the recovered posterior distribution with the true distributions on the Park function for a given Gaussian. 
 The 68\% confidence interval (one sigma) is constructed by taking the central 68\% of the Monte Carlo chain samples and excluding the most extreme 32\% of values i.e. the lowest 16\% and highest 16\% of the distribution. This interval represents the range of values that are most probable given the hierarchical hyperparameters.

 Table~\ref{tbl:distrubiton_error_park} compares the errors in the moments of the inferred distribution with respect to the distribution $P(\theta_{true})$.

\begin{table}[ht]
\centering
\caption{Comparison of inferred hierarchical parameters to their corresponding true values}
 \begin{tabular}{||l c c c c||} 
 \hline
	 & Moment &  Posterior & Uncertainty & Error \\ 
 \hline
     Gaussian  & $\mu$ & 0.5956 & 0.0173 & 0.0035 \\ 
	       & $\sigma$ & 0.0358 & 0.0186 & 0.0136 \\

 \hline
 \end{tabular}
	\label{tbl:distrubiton_error_park}
\end{table}

\subsection{Estimate 2D Distribution using Hierarchical Parameters}

Using the specifications in Table~\ref{tbl:cantilever-specification} the \BC can be applied to the cantilever beam functions, making use of two input dimensions, two parameter dimensions $\theta^{(1)}$ and $\theta^{(2)}$, as well as two outputs $f_D$ and $f_S$ (Equations~\ref{eq:f_d_exp}-\ref{eq:f_s_sim}). In order to help the \GP and bring the output values into the same order of magnitude, the logarithmic values of the outputs are used in the calibration.  

\begin{table}[ht]
\centering
\caption{Specifications used to show the recovery of a given parameter distribution using the hierarchical \BC.}
 \begin{tabular}{|l | c | c | c | c | c| c| c |} 
 \hline
 & \textbf{$N_{MBC}$} & NUTS-MC chain length & \textbf{$P(\theta_{true}^{(i)})$} & \textbf{$|D_{sim}|$} & \textbf{$|D_{exp}|$} & Prior $\theta^{(i)}$ & i \\ 
 \hline
     \textbf{Specifications} & 8 & 450 & $\mathcal{N}(0.6479,0.0255)$ & 400 & 10 & $\mathcal{U}(0.1,1)$ & 1 \\ 
			     & & & $\mathcal{N}(0.6439,0.0169)$ & & 10 & $\mathcal{U}(0.1,1)$ & 2 \\
 \hline
 \end{tabular}
	\label{tbl:cantilever-specification}
\end{table}

\begin{figure}[ht]
    \centering
    \includegraphics[width=0.75\textwidth]{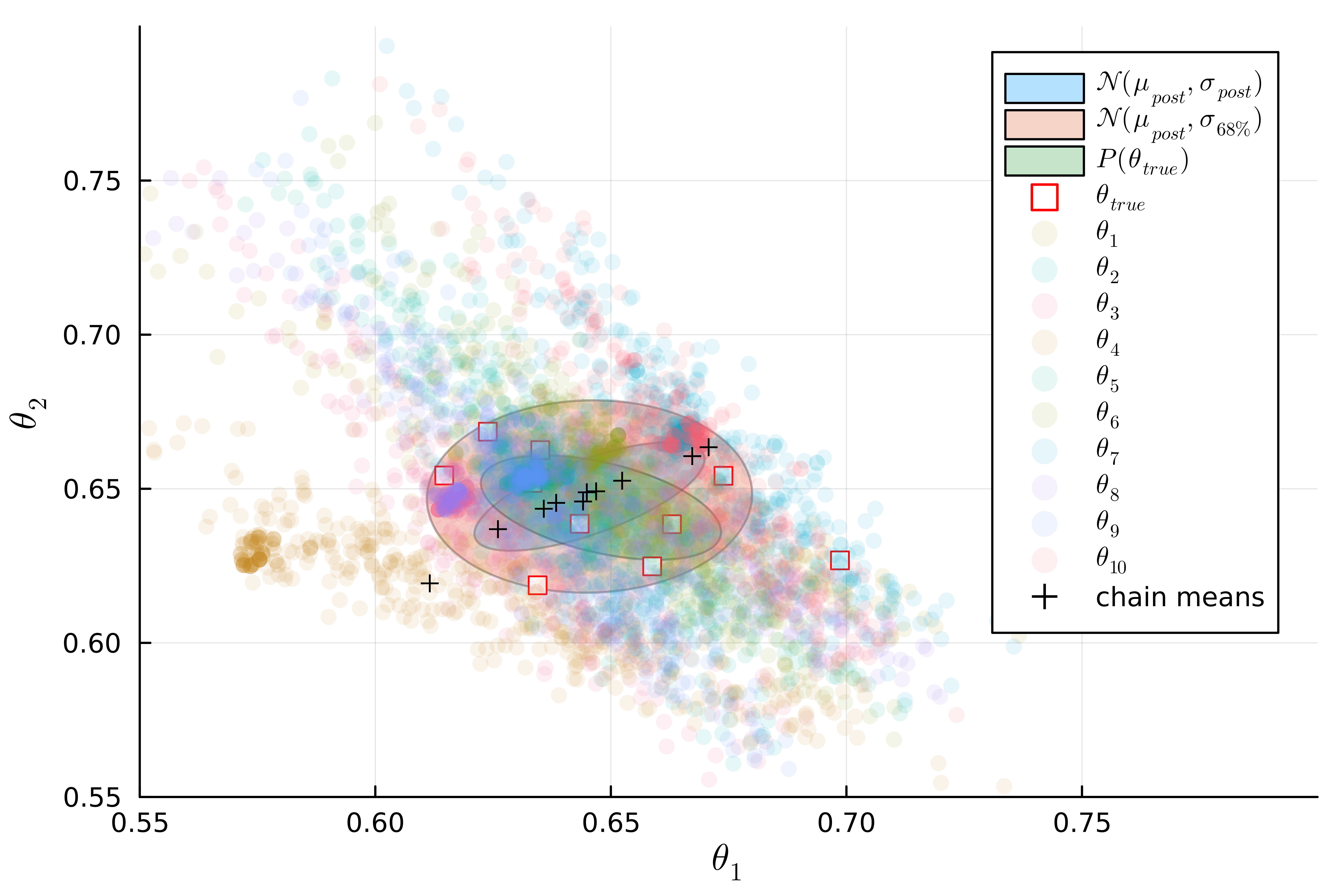}
    \caption{
    Covariance structure in the $\theta$ dimensions comparing the true covariance with the inferred posterior covariance. The notation 
$\mathcal{N}(\mu_{post}, \sigma_{68\%})$
denotes the posterior mean together with the corresponding 68\% interval. The plot also displays the individual sampling chains associated with each synthetic observation, as well as their respective chain-wise means.}

    \label{fig:Covariance_clb}
\end{figure}

\begin{figure}[ht]
    \centering
    \begin{subfigure}[t]{0.5\textwidth}
	\centering
	\includegraphics[width=\textwidth]{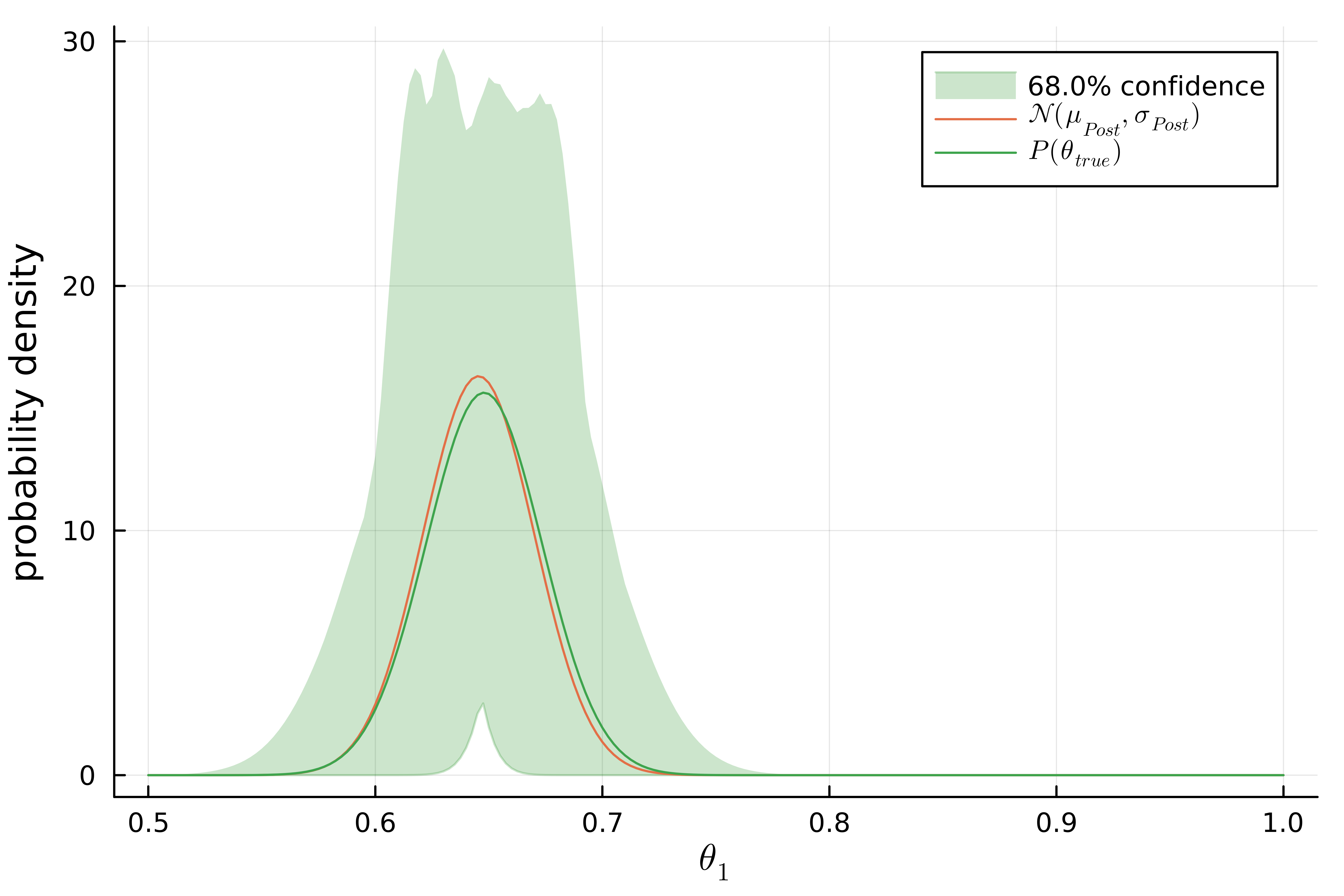}
	\caption{Inferred hierarchical PDF for first $\theta$ dimension}
	\label{fig:pdftheta1clb}
    \end{subfigure}
    \begin{subfigure}[t]{0.5\textwidth}
	\centering
	\includegraphics[width=\textwidth]{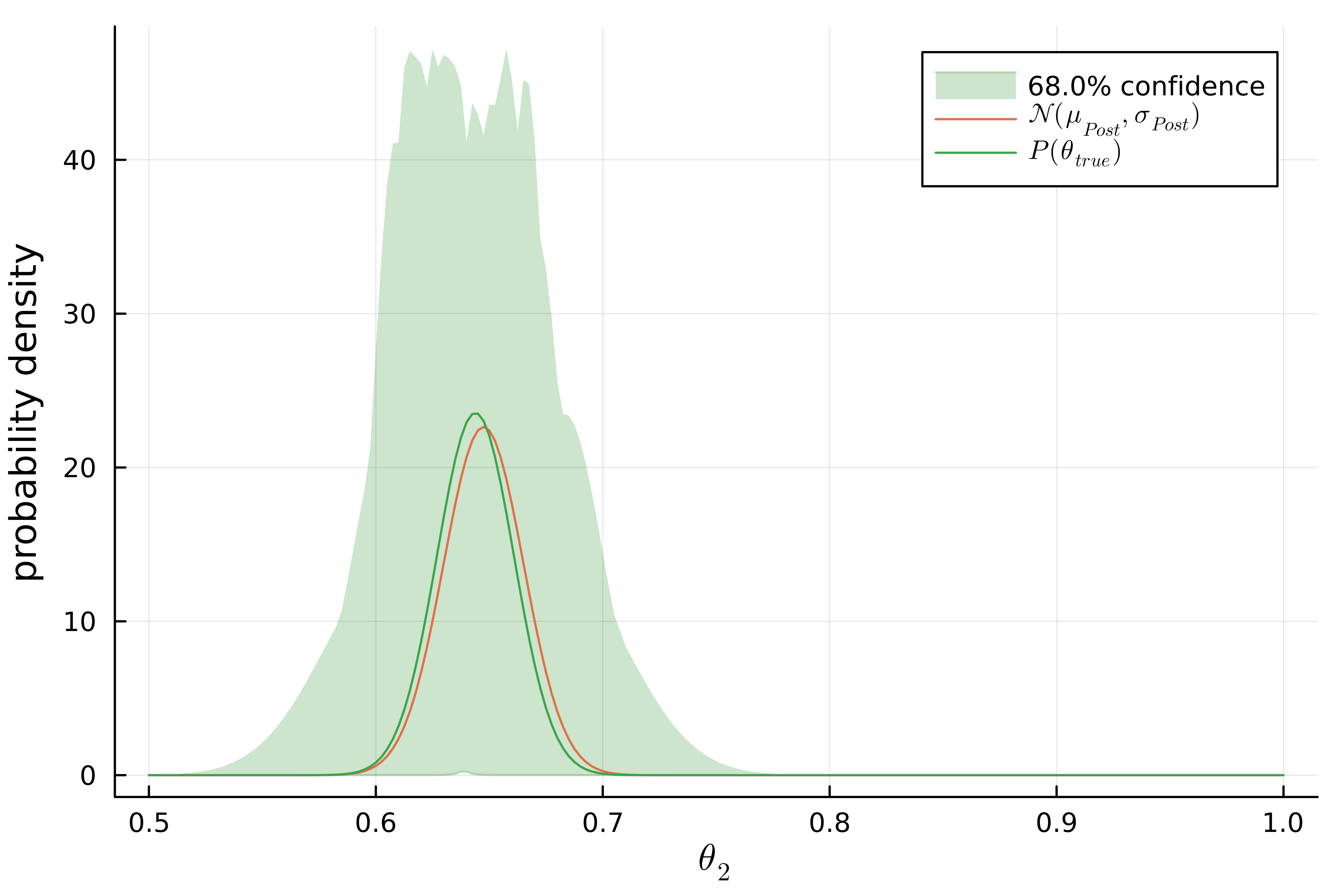}
	\caption{Inferred hierarchical PDF for second $\theta$ dimension}
	\label{fig:pdftheta2clb}
    \end{subfigure}
    \caption{The inferred hierarchical PDF compared with the $P(\theta_{true})$ calculated from the moments of the synthetic observations.}
    \label{fig:pdfsclb}
\end{figure}

Figure~\ref{fig:pdfsclb} shows how for the individual dimension the hierarchical parameter was recovered by the inference compared to the true distribution.\ Figure~\ref{fig:Covariance_clb} shows a covariance plot for both dimensions. The NUTS-MC chains for the individual observations are displayed individually, together with the hierarchical distribution calculated from the $\theta_{true}$ and the inferred hierarchical posterior distribution $\mathcal{N}(\mu_\theta,\sigma_\theta)$. The correlation in the individual observation chains is caused by  the form of the cantilever beam functions.


\subsection{Detailed Runtime and Error Analysis}
To assess the advantages of integrating the \BCM into the \BC framework, this section analyses how error and runtime are influenced by the NUTS-MCMC chain length and the size of the simulation data set. The committee size of \BCM is varied between 1 and 8 members. Since the asymptotic behaviour is more pronounced for a higher number of parameters, this study uses the cantilever beam functions.

\begin{figure}[ht]
    \centering
    \begin{subfigure}[b]{0.5\textwidth}
	\centering
	\includegraphics[width=\textwidth]{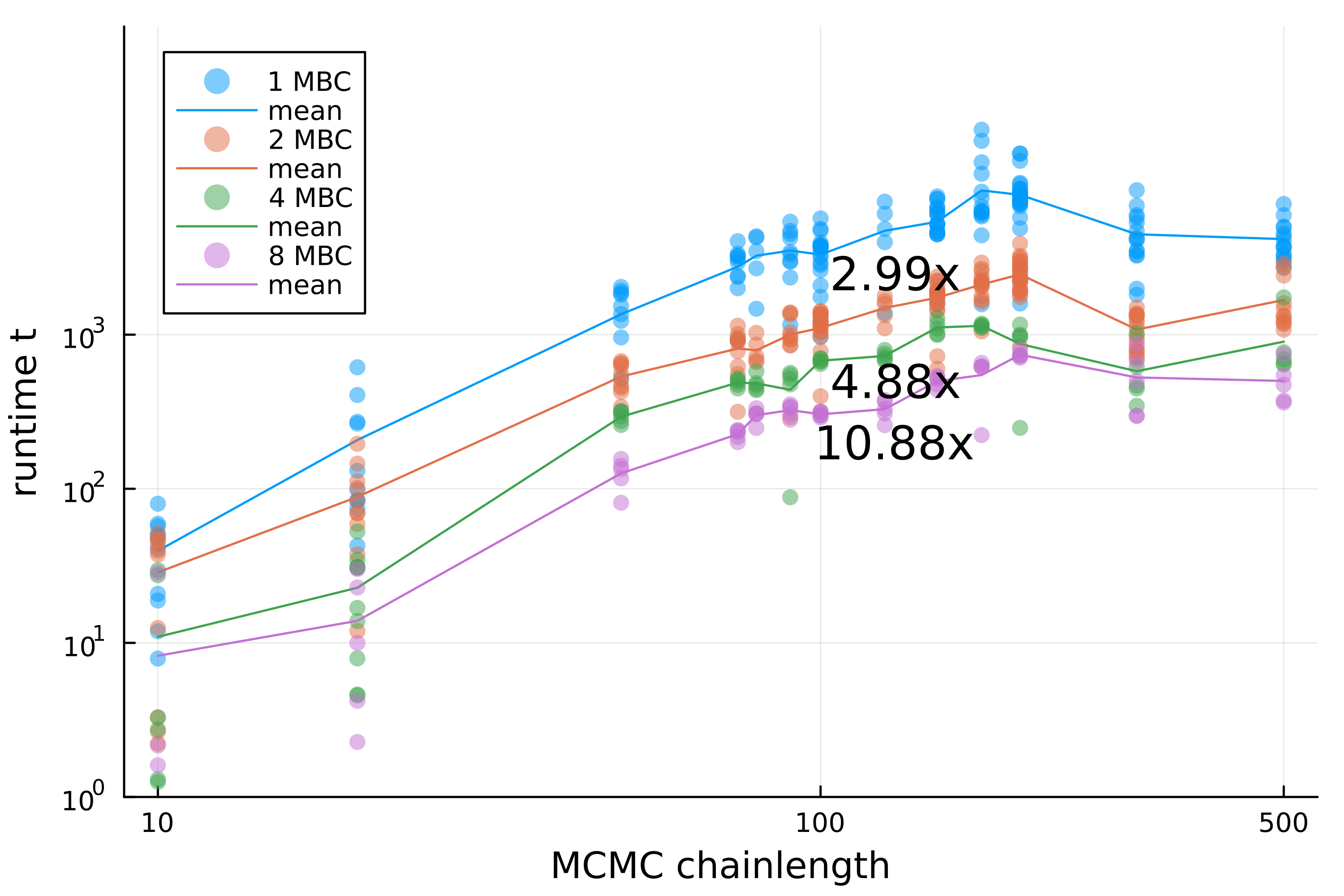}
	\caption{Simulation data set $|D_{sim}| = 100$}
	\label{fig:timebycln100}
    \end{subfigure}
    \begin{subfigure}[b]{0.5\textwidth}
	\centering
	\includegraphics[width=\textwidth]{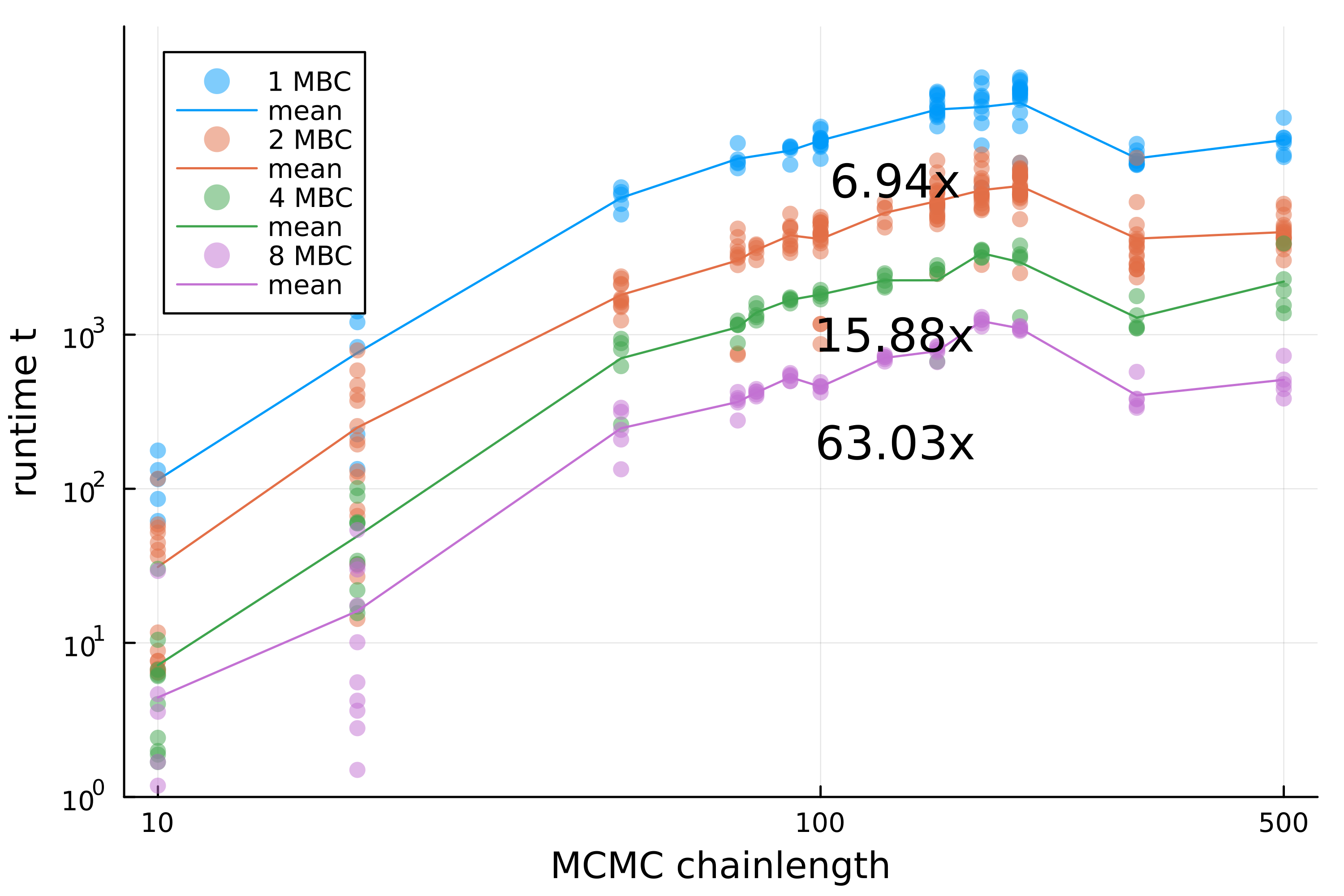}
	\caption{Simulation data set $|D_{sim}| = 200$}
	\label{fig:timebycln200}
    \end{subfigure}
    \caption{Runtime of hierarchical \BC with and without \BCM for two different simulation data set sizes $|D_{sim}|$. The annotations in black are the speed-up over one \MBC at chainlength of 200.}
\end{figure}

Figures~\ref{fig:timebycln100} and~\ref{fig:timebycln200} illustrate the relationship between runtime and MCMC chain length to calibrate using a fixed simulation data set size.

\begin{figure}[ht]
    \centering
    \begin{subfigure}[b]{0.5\textwidth}
	\centering
	\includegraphics[width=0.9\textwidth]{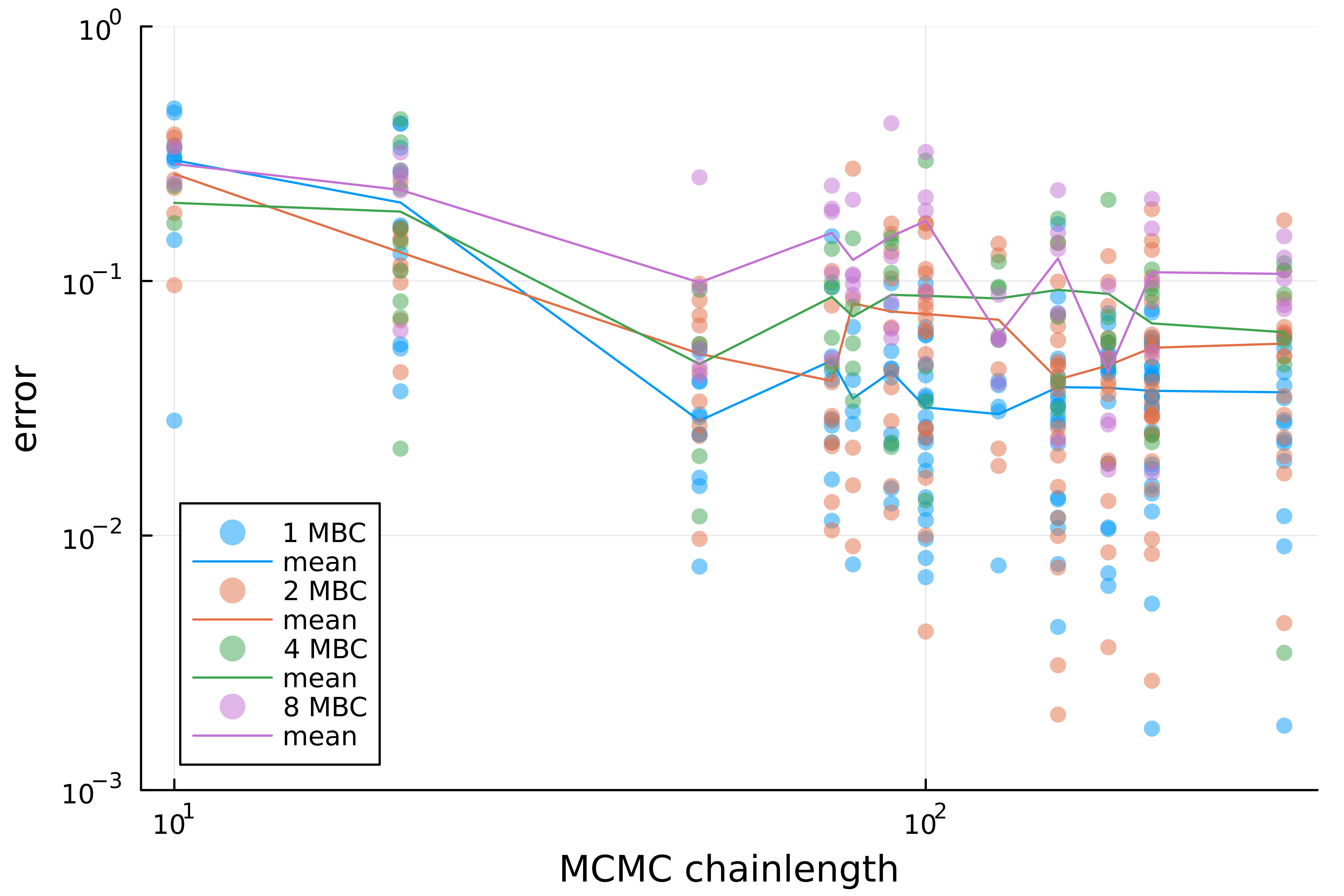}
	\caption{Simulation data set $|D_{sim}| = 100$}
	\label{fig:errorbycln100}
    \end{subfigure}
    \begin{subfigure}[b]{0.5\textwidth}
	\centering
	\includegraphics[width=0.9\textwidth]{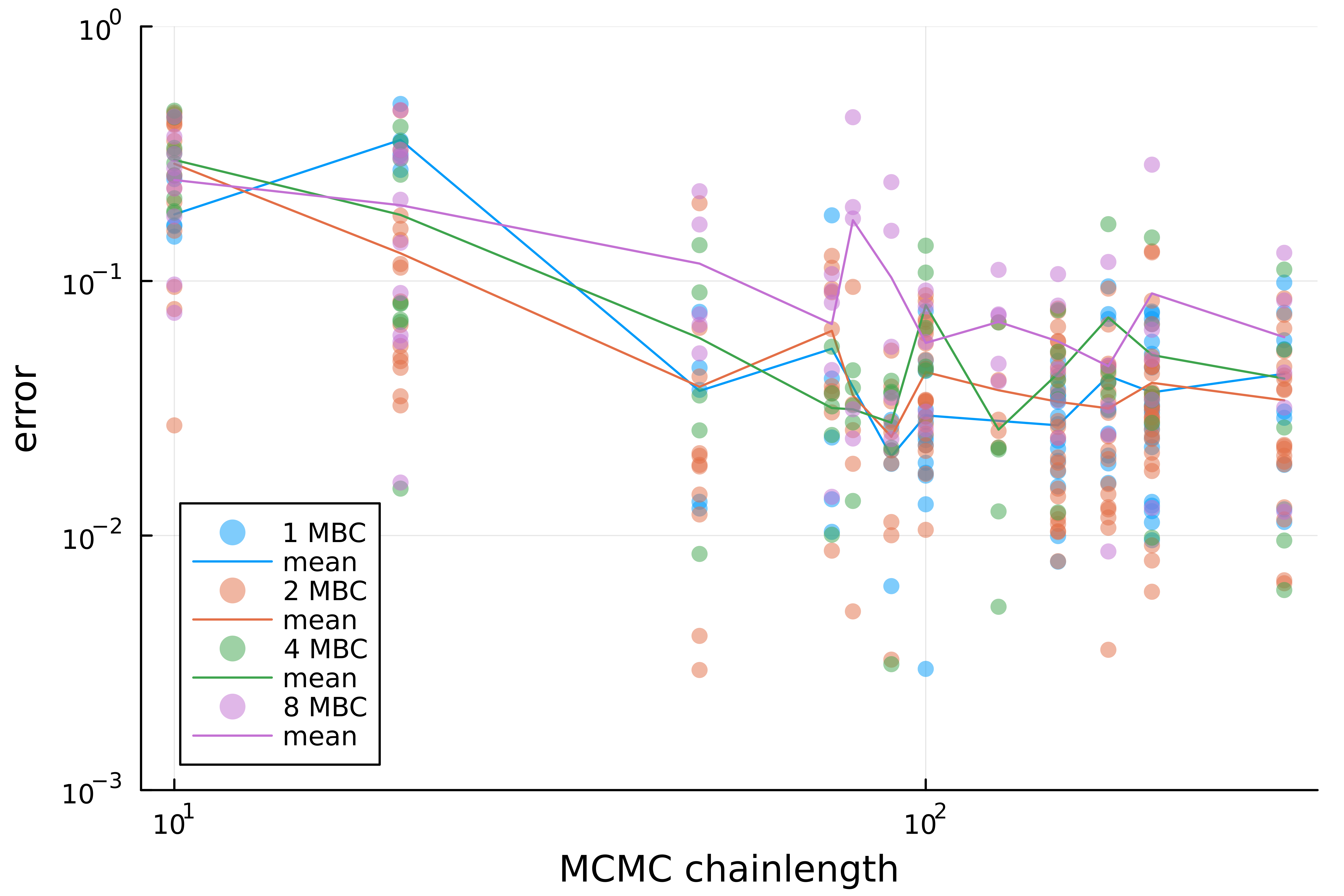}
	\caption{Simulation data set $|D_{sim}| = 200$}
	\label{fig:errorbycln200}
    \end{subfigure}
    \caption{The square root error on the moments of hierarchical \BC with and without \BCM dependent on the sample chain length for two different simulation data set sizes $|D_{sim}|$.}
    \label{fig:errorbycl}
\end{figure}

Figures~\ref{fig:errorbycl} show how far the expected mean square error on the moments converges with longer sample chains.  

The convergence of chains was not explicitly evaluated in this analysis; however, as indicated in Figure~\ref{fig:errorbycl}, a minimum chain length of approximately 100 samples is typically required to achieve sufficient convergence for this problem.

\begin{figure}[ht]
    \centering
    \begin{subfigure}[b]{0.5\textwidth}
	\centering
    \includegraphics[width=0.9\textwidth]{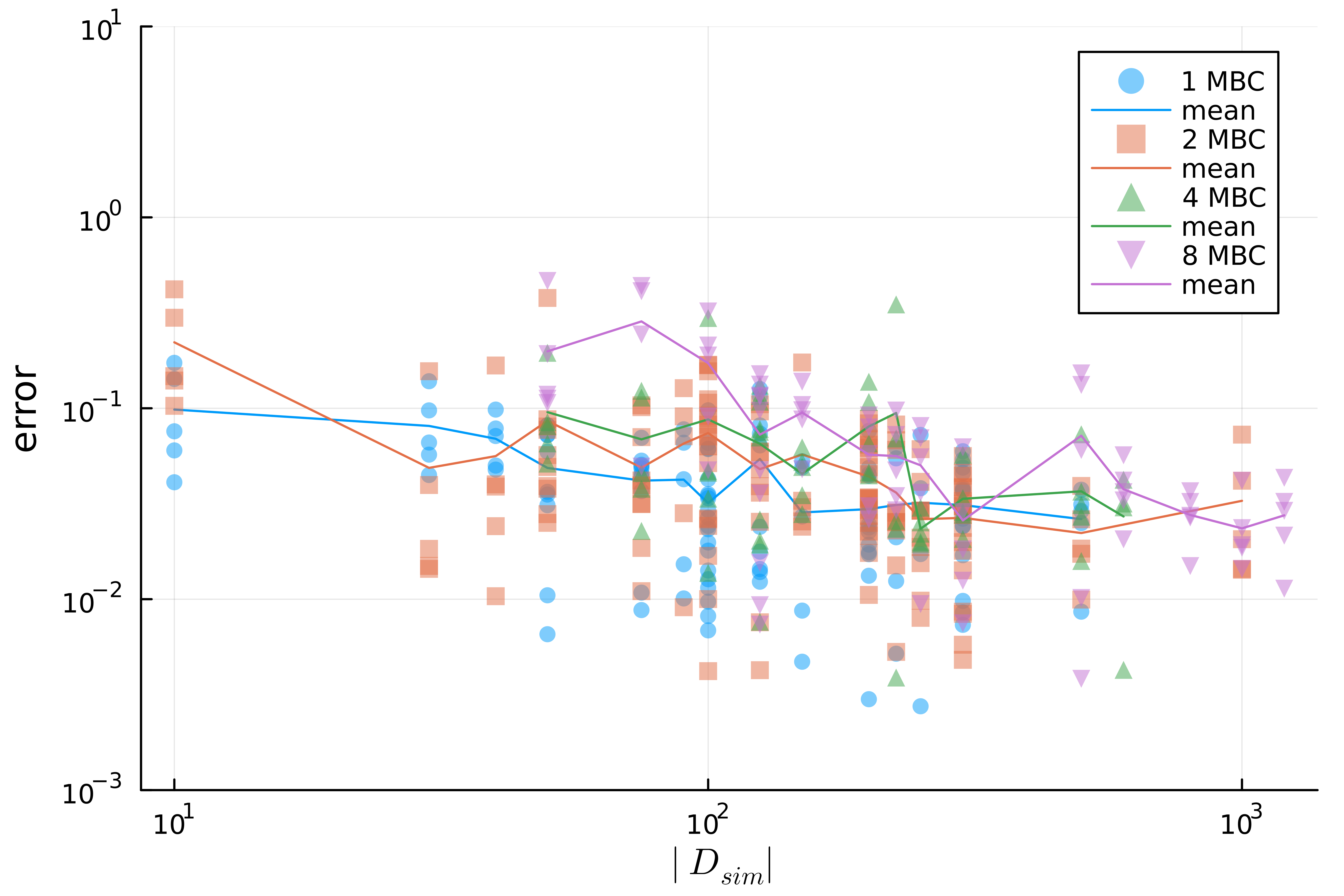}
    \caption{Different simulation data set sizes}
    \label{fig:errorbydata}
    \end{subfigure}
    \begin{subfigure}[b]{0.5\textwidth}
	\centering
    \includegraphics[width=0.9\textwidth]{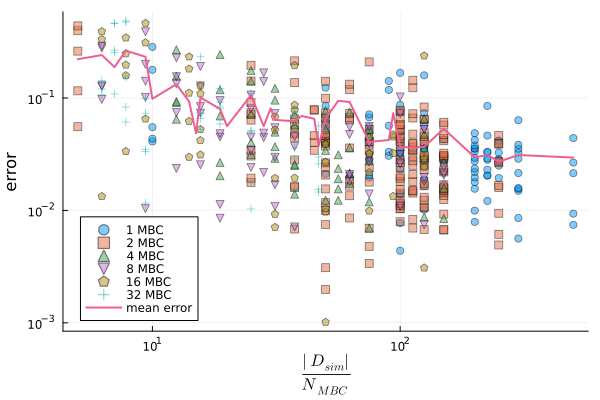}
    \caption{Different data ratios ($\frac{|D_{sim}|}{N_{MBC}}$)}
    \label{fig:errorbydataratio}
\end{subfigure}
\caption{The square root error on the moments of the \BC at sample chain length 100 against total data set size and different data ratios.}
\end{figure}

Figure~\ref{fig:errorbydata} demonstrates how larger simulation data sets influence the error. The error is again measured as the mean square error on the moments. The effect of too small data sets per individual member of the Bayesian Committee is illustrated by Figure~\ref{fig:errorbydataratio}. In this plot, the mean square error on the moments is plotted against the data ratios $\frac{|D_{sim}|}{N_{MBC}}$ and shows that only for less than 20 data points per \MBC, the splitting leads to a larger error in the inferred calibration parameter.

\begin{figure}[ht]
    \centering
    \includegraphics[width=0.7\textwidth]{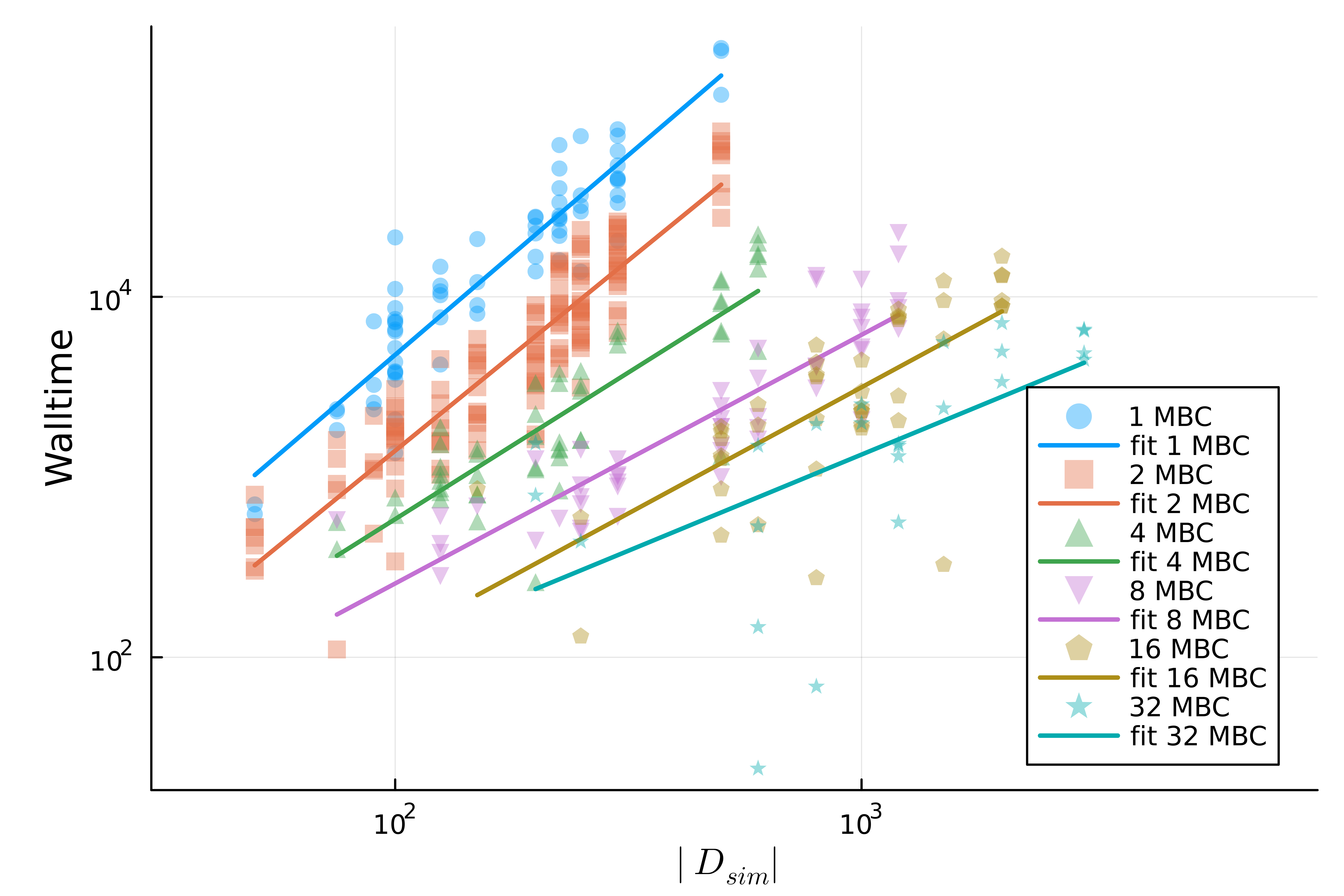}
    \caption{The runtime of the \BC with an error $\epsilon \le 0.01$ for different simulation data set sizes with sample chain lengths between 50 and 3000.}
    \label{fig:neccessaryruntime}
\end{figure}

As a more practical result, Figure~\ref{fig:neccessaryruntime} summarises the interfering effects on the runtime and error of the objectives shown by the other plots. In this study of the runtime only those sampling chains that achieved a squared error of the moments $\epsilon \le 0.01$ are included, while different lengths of sampling chains are pooled. The assumption is that the aim is an a priori level of accuracy and that the necessary simulation data set size for a reasonable resolution depends on the specific problem function. This data size can be estimated by how well the pre-optimised \GP for $\eta_{sim}$ can predict the simulation data.

\clearpage 
\pagebreak 

\subsection{Argonne Wakefield Accelerator Beam Test Facility} 
For the AWA electron gun application described in Section~\ref{s:awa}, we demonstrate the methodology using simulated experimental data generated from OPAL simulations.

In this setup, the gun injection amplitude $V_{gun}$ is the parameter to be inferred, while $\phi_{gun}$ and the solenoid currents $I_{foc}$, $I_{match}$, and $I_{buck}$ are treated as known inputs. As measurement output, the beam energy at the location of the first monitor is used. 

The simulation data generated $D_{sim}$ and the experiment data $D_{exp}$ were drawn from uniform distributions $\mathcal{U}(a,b)$ in the ranges given in Table~\ref{tbl:AWAdatasets}. For the simulation also the parameter $V_{gun}$ is varied within its uniform prior, while for the synthetic observation $V_{gun} \sim \mathcal{N}(55MV,2MV)$.

\begin{table}[h!]

\caption{Data set parameters and inputs used for the AWA gun example.}
\centering
 \begin{tabular}{||l c c c ||} 
 \hline
	 & $D_{sim}$ &  $D_{exp}$ &  \\ 
 \hline
     $ N = | D |$ & 2400 & 15 & \\
 \hline
     $V_{gun}$ & $\mathcal{U}(50,70)$ &  $\mathcal{N}(55,2)$ & MV \\ 
 \hline
     $\phi_{gun}$ & $\mathcal{U}(0,50)$ &  $\mathcal{U}(0,50)$ & deg \\ 
     $I_{foc}$/$I_{match}$ & $\mathcal{U}(160,260)$ & $\mathcal{U}(160,260)$ & A \\ 
     $I_{buck}$ & $\mathcal{U}(430,570)$ & $\mathcal{U}(430,570)$ & A \\
 \hline
 \end{tabular}
	
	\label{tbl:AWAdatasets}
\end{table}

\begin{figure}[h!]
    \centering
    \begin{subfigure}[b]{0.5\textwidth}
	\centering
	\includegraphics[width=\textwidth]{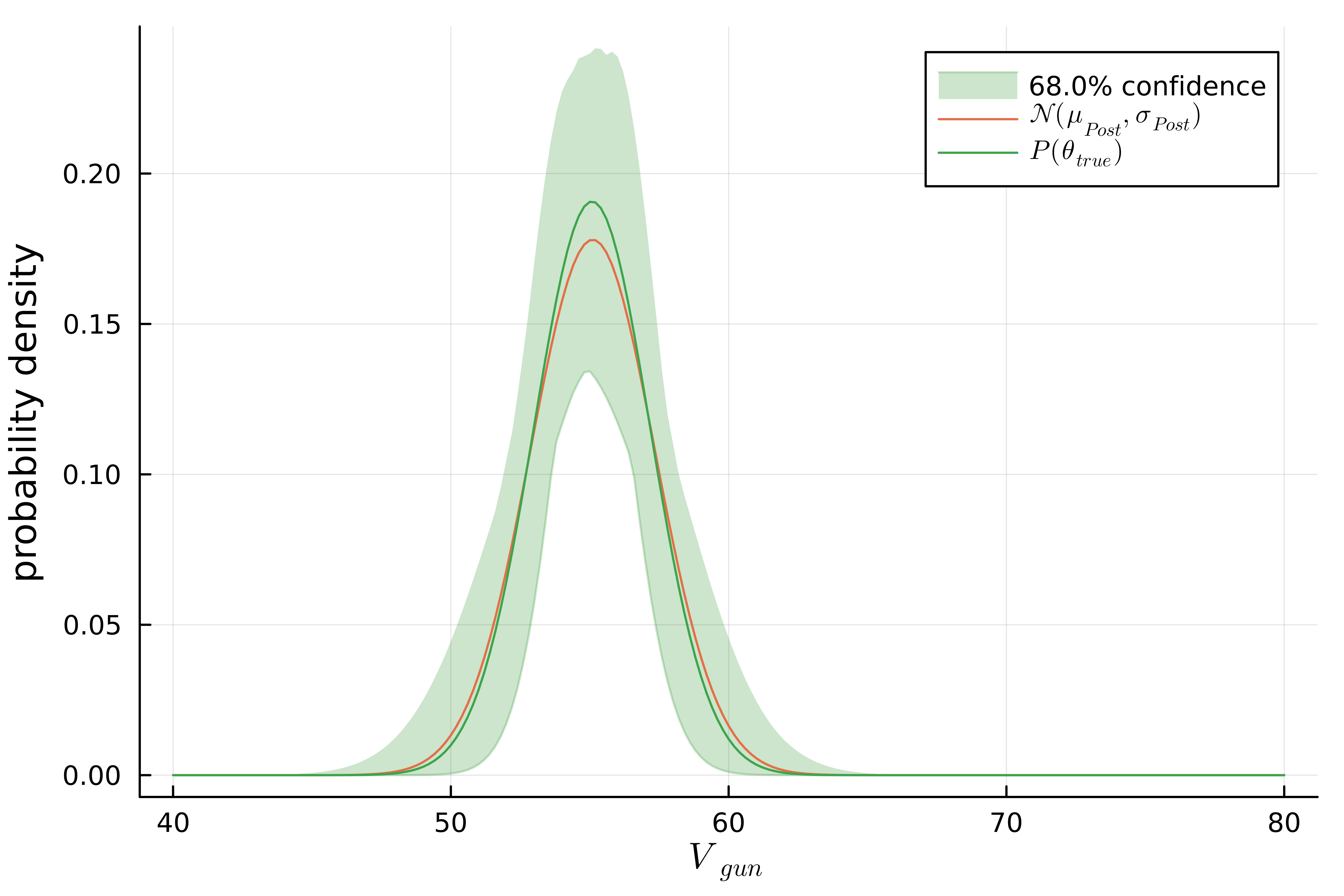}
	\caption{Inferred hierarchical gun injection amplitude PDF \\ \hspace{1em} }
	\label{fig:pdfawa}
    \end{subfigure}
    \begin{subfigure}[b]{0.5\textwidth}
	\centering
	\includegraphics[width=\textwidth]{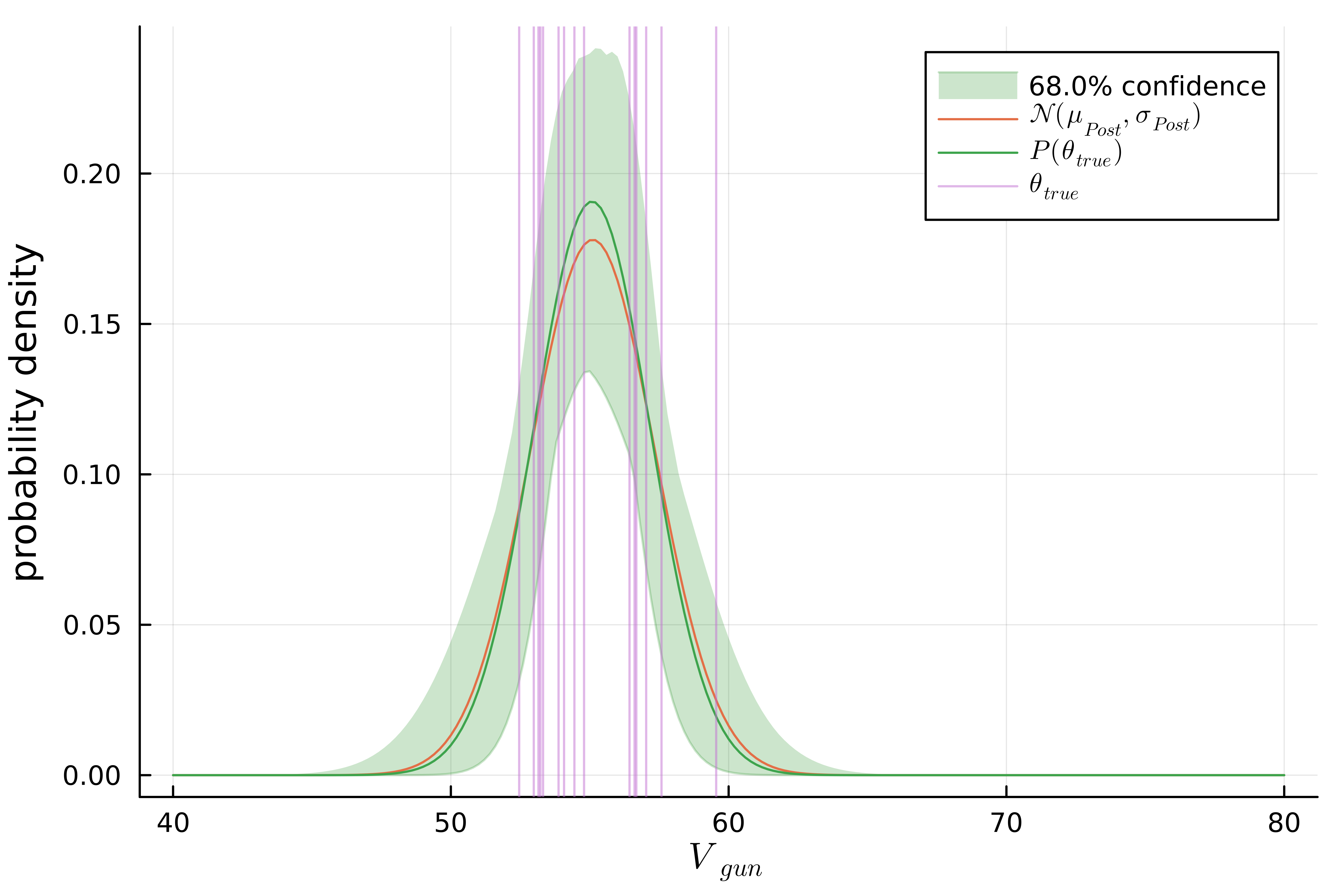}
	\caption{Inferred hierarchical gun injection amplitude PDF with data histogram}
	\label{fig:pdfawawithhist}
    \end{subfigure}
    \caption{The inferred hierarchical gun injection amplitude PDF compared with the $P(\theta_{true})$ calculated from the moments of the simulated measurements. }
    \label{fig:awapdfs}
\end{figure}

\begin{table}[ht]
\caption{The inferred $\vartheta$ gun injection amplitudes for the 15 synthetic observations as well as the hierarchical parameters. The column $D_{exp}$ states the actual value in the observation data set, while the columns $\mathbb{E}$ and $\Sigma$ show the mean and standard deviation calibrated from the Markov chain. The $D_{exp}$ entries for $\mu_{post}$ and $\sigma_{post}$ are calculated from the 15 observations.}
\centering
 \begin{tabular}{||l c c c ||} 
 \hline
	$\theta_i$ & $D_{exp}$ & $\mathbb{E}[ \cdot ]$ & $\Sigma[ \cdot ]$\\ 
 \hline
     
    1 & 53.1532  & 53.2286 & 0.0357 \\
2 & 57.0313 & 54.452  & 0.0274 \\
3& 56.613  & 57.5996 & 0.0286\\ 
4& 54.7952 & 54.7703 & 0.0274\\
5& 57.5829 & 53.8301 & 0.0284\\
6& 56.4317 & 59.4781 & 0.0312\\
7& 53.2223 & 54.0388 & 0.0398\\
8& 53.8767 & 53.1315 & 0.0413\\
9& 54.0751 & 56.2681 & 0.0323\\
10 &54.4473 & 53.2797 & 0.0404\\
11 &52.9847 & 56.431  & 0.0339\\
12 &59.5515 & 56.7266 & 0.0287\\
13 &52.4592 & 56.9984 & 0.0297\\
14 &56.6766 & 53.1593 & 0.0292\\
15 &53.3191 & 52.734  & 0.0381\\
\hline
     $\mu_{post} $ & 55.0813 &    55.1060  &   0.5911 \\
     $\sigma_{post}$ & 2.0917 &      2.2410 &    0.5249 \\
 \hline
 \end{tabular}
	\label{tbl:AWAinferredGun}
\end{table}

 Table~\ref{tbl:AWAinferredGun} shows how the inferred parameter vector $\vartheta$ (\Eq~\ref{eq:vartheta}) of individual $\theta_i = V_{gun,i}$, and the hierarchical parameters $\mu_{post}$ and $\sigma_{post}$ compare to the corresponding true values in the data set $D_{exp}$. It shows that the individual parameters are recovered as expected and that the true value lies within the standard deviation of the corresponding NUTS-MC chain. The relation of the observation values with the inferred hierarchical distribution can be seen in Figure~\ref{fig:pdfawawithhist}.

\section{Discussion}

The Hierarchical Bayesian Calibration works as expected and is successful in recovering the distribution $P(\theta_{true})$ drawn as ground truth for the 10 observations (cf.\ Section 3.3). This increases the dimensions of the parameter vector that need to be inferred but has an edge over the ordinary KOH Bayesian Calibration, which requires all experiments to use the same parameter settings. This advantage of the Hierarchical Parameters is particularly useful even when the individual parameter distributions are not Gaussian or unimodal. The successful calibration is evident not only in Figure~\ref{fig:gaussian_pdf_theta_park} or~\ref{fig:Covariance_clb}, but also in the convergence of the error in Figure~\ref{fig:errorbycl} and~\ref{fig:errorbydata} for longer NUTS chains and larger data sets. This further corroborates the accuracy and reliability of the implementation. 

Figure~\ref{fig:kohOdataset} clearly shows the advantage of using Bayesian Committee Machines in the context of \BC. While not splitting the data sets leads to a cubic growth of the runtime in the number of data points, applying \BCM significantly reduces the runtime. In practice, for a data set of size $|D_{sim}| = 150$, we achieved a speed-up of 15.86 for eight \MBCS.

Comparing the error of models with two to eight \MBCS to the version without Bayesian Committee (1 \MBC), one sees that the versions with Bayesian Committee are not expected to converge any slower. This is clear from Figures~\ref{fig:errorbycl} showing the expected error of multiple runs over the Markov chain length. This also shows how a certain chain length is necessary for the calibration to converge. 

Figures~\ref{fig:errorbydata} ~ and~\ref{fig:errorbydataratio} show the effect of splitting up the data sets on the error of the moments of the inferred parameters.  As Figure~\ref{fig:errorbydataratio} shows, a very small number of data points per \MBC leads to a larger error. As a result, the examples with four or eight \MBCS start to have worse mean errors for small data sets, but the models do catch up to the same error for data ratios $\frac{|D_{sim}|}{N_{MBC}} > 20$. 
This is an argument against too many \MBCS even if enough processors are available, but this effect disappears when each \MBC gets assigned a reasonable number of data points.  
Figure~\ref{fig:errorbydata} also shows that even for such a simple problem, the error decreases with a larger number of simulation data points.

These error plots show, for our examples, that once each \MBC contains a sufficient number of simulation results,
the error on the inferred moments is of similar magnitude as in the 1 \MBC case (i.e. no \BCM approximation). We conclude that here the \BCM approximation is accurate and samples the same target distribution. However, without known bounds on the approximation error, this cannot be generalised to all problems, and the effect on the exact target distribution for other types of problems needs to be further investigated. 

With regard to the points mentioned in Section~\ref{ss:valofBCM}, the use of \BCM to estimate the likelihood of parameters of multiple experiments at once, as is the case in this hierarchical framework, is supported by theory, more so than its use in single-point prediction.

The underestimation of the uncertainty that can occur in prediction with the \BCM approximation is also less problematic in this framework. For one, if the support of the simulation data is similar to or contains the support of the parameter prior, the BCM is not evaluated far away from the data points. On a higher level, the hierarchical \BC increases the variance to capture the variance between the experiments. Only when the uncertainty in the simulation data is larger than the between experiment variance, this could have an effect on the inferred second moment.

From those results one can see that, first, the amount of simulation data necessary depends on the problem; the GP prediction error can be decreased with a larger simulation data set. Second, the MCMC chain must be sufficiently long to adequately explore the posterior and for the Monte Carlo-based errors to be sufficiently small. This MCMC run length depends mainly on the problem and the posterior distribution. Hence, both the necessary chain length and the amount of necessary data can be seen as problem dependent. 

Therefore, the addition of the Bayesian Committee Machine to Bayesian Calibration is beneficial to the runtime. The problem requires a certain MCMC chain length and a certain number of data points. Using twice the number of \MBCS one can also double the number of simulation data points and still achieve at most the same expected runtime. 
As can be seen, for example, in Figure~\ref{fig:neccessaryruntime} it is clearly beneficial to use the \BCM to calibrate with large data sets. The main thing that needs to be avoided is a too low data ratio.

For example, given that the number of necessary data points is around $|D_{sim}| = 100 $ to $200$ and the NUTS-MCMC chain length around 100. For this example, one can read in Figure~\ref{fig:timebycln100} that this places us in the region between 50 and 60 minutes without the Bayesian Committee Machine, while with eight \MBCS the runtime stays around 5 minutes. For a problem that needs 200 data points, the ordinary version has an expected runtime of hours, while eight \MBCS still stay below the ten minute mark. A notable difference for an operator in the experimental booth. 

An aspect of the \BCM approximation not explored in this study involves using domain expertise to perform domain-specific decomposition of simulation data sets, rather than employing the random splits utilised here. Such domain-driven decomposition could prove advantageous in scenarios where the underlying calibration function exhibits highly dynamic responses concerning particular inputs or parameters.

Another potential enhancement mentioned in Section~\ref{ss:BCM} is the integration of \BCM with optimal subset selection or decomposition methodologies. After partitioning the data set into subsets, the Bayesian Committee Machine employs conventional Gaussian Processes, which can be further approximated and accelerated using the techniques outlined in the introduction. These approximation methods, characterised by asymptotically superlinear computational complexities, would benefit substantially from reduced subset sizes and inherent parallelisation capabilities.

It is important to acknowledge explicitly that the method introduced here contains inherent bias due to prescribing the target distribution explicitly. Analogous to the careful selection of priors, the parametric distribution within Hierarchical Bayesian Calibration should be chosen thoughtfully, and its influence on results must be thoroughly analysed for sensitivity. Employing a general parametrised distribution or a mixture model would likely mitigate such biases, offering the most versatile solution. Although the current framework can accommodate other hierarchical distribution forms, investigating these was outside the scope of the present study.

\section{Conclusion}

This work presents a scalable implementation of Hierarchical Bayesian Calibration that integrates the Bayesian Committee Machine to address computational bottlenecks arising from large simulation data sets. The approach extends the Kennedy–O’Hagan framework by combining experiment-level calibration parameters with a latent hierarchical structure, enabling partial pooling and inference of these separate, but related parameters. This formulation is particularly relevant in applications where the calibration parameters vary between repeated or evolving experimental conditions.

Our results demonstrate that incorporating the BCM yields substantial computational gains while preserving inferential accuracy. The method achieves favourable computational scaling, reduces the effective cost of Markov chain Monte Carlo sampling, and enables the use of high-fidelity simulators in calibration tasks that would otherwise be computationally prohibitive. Because the BCM partitions the data rather than relying on a data set-specific low-rank or inducing-point approximation, it is less prone to biases such approximations might introduce.

Through controlled benchmark problems and a realistic accelerator case study, the approach is shown to recover parameter distributions accurately, quantify uncertainty reliably, and scale effectively with available simulation data.

A further practical component of this work is the implementation of the framework using the Julia ecosystem—specifically Turing.jl for probabilistic modelling and AdvancedHMC.jl for gradient-based sampling. Automatic differentiation enables efficient evaluation of high-dimensional log-posterior gradients, eliminating the need for manual derivations and allowing flexible model specification without imposing restrictive assumptions. Combined with the BCM decomposition strategy, this enables computationally tractable posterior sampling even when the calibration model relies on thousands of simulation evaluations.

This novel combination of Hierarchical Bayesian Calibration with the Bayesian Committee Machine and the No-U-Turn Sampler is validated against the Park and cantilever beam benchmark functions, as well as a real-world accelerator example from the Argonne Wakefield Accelerator Beam Test Facility. The improvements in computational performance are demonstrated in scaling studies. 

Overall, the proposed framework provides a statistically coherent, computationally efficient, and practically deployable solution for hierarchical calibration in data-constrained, simulation-driven experimental science. \\[1cm]
 
\noindent {\bf Declaration:} The authors confirm that there are no relevant financial or non-financial competing interests to report.

\noindent {\bf Data availability: } Raw data were generated using OPAL~\cite{adelmannOPALVersatileTool2019} and sampling the benchmark functions defined in Section 3.1. Derived data supporting the findings of this study are available from the corresponding author AA on request.

\newpage

\appendix

\section{Automatic Relevance Determination in Multivariate Kernel}
\label{ap:adr}

Following \cite{williamsGaussianProcessesRegression1995}, the kernel can easily be extended to multiple or vector valued inputs by choosing a suitable distance function $r(\bm{x},\bm{x'})$. This way the kernel processes the inputs $x$ and parameters $\theta$ at the same time. In such contexts it is helpful to use independent length scales for individual dimensions as this allows the kernel to adapt not only to different input scales but also to weigh the relevance of each input dimension. 

\begin{align}
	\label{eq:ardgp}
	k_{\lambda,\bm{l}}(\bm{x},\bm{x}') &= \lambda k\left( \langle \bm{l}, \begin{pmatrix} r(x_1,x_1')\\ r(x_2,x_2') \\ \vdots \end{pmatrix} \rangle\right) 
\end{align}
	
For such an Automatic Relevance Determination (\Eq~\ref{eq:ardgp}), $\bm{l}$ is a vector of inverse length scales. The distance $r(x,x')$ is applied element-wise to the input dimensions, and both are combined by a scalar product.

For a map from inputs $x \in \mathbb{R}^n$ to outputs $z \in \mathbb{R}$, the full Mat\'ern-$\sfrac{3}{2}$ kernel function with the rescaling hyperparameter $\lambda$ and inverse length scale $l$ is given by \Eq~\ref{eq:covfullx}.

\begin{equation}
	k_{\lambda,\bm{l}}(\bm{x},\bm{x'}) = \lambda \left( 1 + \sqrt{3 \sum_i^n l_i^2 (x_i -x'_i)^2} \right)\exp\left( - \sqrt{3 \sum_i^n l_i^2 (x_i -x'_i)^2 } \right)   
\label{eq:covfullx}
\end{equation}

For a map that includes the parameter $\theta \in \mathbb{R}$ on the input side, the full kernel function is given by \Eq~\ref{eq:covfullxtheta}.

\begin{equation}
	k_{\lambda,\bm{l}}(\bm{x},\theta,\bm{x'},\theta') = \lambda \left( 1 + \sqrt{3 \sum_i^n l_i^2 (x_i -x'_i)^2 + l_{n+1}^2(\theta-\theta')^2 }\right)\exp\left( - \sqrt{3 \sum_i^n l_i^2 (x_i -x'_i)^2 + l_{n+1}^2 (\theta-\theta')^2 }\right)   
\label{eq:covfullxtheta}
\end{equation}

\section{Predictions and Probabilities}
\label{ap:pred}

We can think of the conditional mean of a \GP as a surrogate for a function or simulation. As such it can be used to predict an output value of the modelled function for specific input $x^*$. More generally, we can compute the posterior distribution of any prediction after conditioning on observations.  The detailed derivations can be found in Gaussian Processes for Machine Learning~\cite{rasmussenGaussianProcessesMachine2005}.

For the prediction of the simulator $\eta$ modelled as defined in \Eq~\ref{eq:etagp} the prediction at a point $(x^*,\theta^*)$ is given by \Eq~\ref{eq:prediction}. \Eq~\ref{eq:predictionvar} describes the associated variance.

\begin{align}
\eta_{\lambda_{\eta},\bm{l}_{\eta}} &\sim \GP\left(\bm{0}, \Sigma_{\sfrac{3}{2},\lambda,l}\left(\begin{pmatrix}X_{sim} & \bm{\theta}_{sim}\end{pmatrix}\right) | \bm{z}_{sim} \right)\label{eq:etagp}
\end{align}

\begin{align}
	\bar{\eta}_{\lambda_{\eta},\bm{l}_{\eta}}(x^*,\theta^*,\sigma) &= \begin{bmatrix} cov_{\lambda_{\eta},l_{\eta}}\left((X_{sim,1},\theta_1),(x^*,\theta^*) \right) 
		\\
		\vdots 
		\\
		cov_{\lambda_{\eta},l_{\eta}}\left((X_{sim,N_{sim}},\theta_{N_{sim}}),(x^*,\theta^*) \right) 
		\end{bmatrix}^T \left( \Sigma_{\sfrac{3}{2},\lambda,l}\left(\begin{pmatrix}X_{sim} & \bm{\theta}_{sim}\end{pmatrix}\right) + \sigma^2 I \right)^{-1} \bm{z}_{sim} \label{eq:prediction}
\end{align}

\begin{align}
\mathbb{V}[\bar{\eta}_{\lambda_{\eta},\bm{l}_{\eta}}(x^*,\theta^*,\sigma) ] = 
cov(x^*,\theta^*,x^*,\theta^*) - \bm{k_*}^T \left( \Sigma_{\sfrac{3}{2},\lambda,l}\left(\begin{pmatrix}X_{sim} & \bm{\theta}_{sim}\end{pmatrix}\right) + \sigma^2 I \right)^{-1} \bm{k_*}
		\label{eq:predictionvar}
\end{align}
\begin{align}
	\text{where }\bm{k_*} = \begin{bmatrix} cov_{\lambda_{\eta},l_{\eta}}\left((X_{sim,1},\theta_1),(x^*,\theta^*) \right) 
		\\
		\vdots 
		\\
		cov_{\lambda_{\eta},l_{\eta}}\left((X_{sim,N_{sim}},\theta_{N_{sim}}),(x^*,\theta^*) \right) 
		\end{bmatrix} \nonumber
		\end{align}

Further, a \GP can be used to calculate the likelihood or log-likelihood (\Eq~\ref{eq:logpdf}) of an output $\bm{y}$ input $\bm{X}$ pair. 

\begin{align}
    \label{eq:logpdf}
	\log p(\bm{y} | \bm{X}) = - \frac{1}{2} \bm{y}^T \left(\Sigma(X,X) + \sigma^2 I\right)^{-1} \bm{y} - \frac{1}{2} \log | \Sigma(X,X) + \sigma^2 I | - \frac{n}{2} \log 2 \pi \\
		\text{where } n = dim(\bm{y}) \nonumber	
\end{align}

By setting

\begin{align}
	\bm{y} &= \begin{bmatrix} z_{sim} \\ z* \end{bmatrix} \text{ and } \\
	\bm{X} &= \begin{bmatrix} X_{sim} & \theta_{sim}  \\ x^* & \theta^* \end{bmatrix}
\end{align}

the \Eq~\ref{eq:logpdf} gives the log-likelihood that ($x^*, \theta^*, z^*$) are the input and output of the same function as ($X_{sim},\theta_{sim},z_{sim}$).

\section{Detailed \texorpdfstring{$\Sigma$}{Sigma}}
\label{ap:sigma}

This shows the detailed construction of the covariance matrix $\Sigma_{\eta,sim}$~(\Eq~\ref{eq:sigmaetasim}).
\begin{align}
	\label{eq:sigmaetasim}
	\Sigma_{\eta,sim} &= {\begin{bmatrix} cov_{\lambda_{\eta},l_{\eta}}\left((X_{sim,1},\bm{\theta}_{sim,1}),(X_{sim,1},\bm{\theta}_{sim,1}) \right) 
		& \cdots 
		&  cov_{\lambda_{\eta},l_{\eta}}\left((X_{sim,1},\bm{\theta}_{sim,1})),(X_{sim,N_{sim}},\bm{\theta}_{sim,N_{sim}}) \right) \\
		cov_{\lambda_{\eta},l_{\eta}}\left((X_{sim,2},\bm{\theta}_{sim,2}),(X_{sim,1},\bm{\theta}_{sim,1}) \right) 
		& \cdots 
		&  cov_{\lambda_{\eta},l_{\eta}}\left((X_{sim,2},\bm{\theta}_{sim,2}) ),(X_{sim,N_{sim}},\bm{\theta}_{sim,N_{sim}}) \right) \\
		\vdots 
		& \ddots & \vdots \\
		cov_{\lambda_{\eta},l_{\eta}}\left((X_{sim,N_{sim}},\bm{\theta}_{sim,N_{sim}}),(X_{sim,1},\bm{\theta}_{sim,1}) \right) 
		& \cdots 
		&  cov_{\lambda_{\eta},l_{\eta}}\left((X_{sim,N_{sim}},\bm{\theta}_{sim,N_{sim}}),(X_{sim,N_{sim}},\bm{\theta}_{sim,N_{sim}}) \right) 
	\end{bmatrix}	}
\end{align}

In a sampling process, the \GP is applied to the inputs $X_{exp}$ in the experimental data $D_{exp}$. Since the experimental data is missing the parameter $\theta$, a trial parameter $\theta^*$ needs to be supplied for the evaluation of the covariance matrix~\Eq~\ref{eq:sigmaetaobs}. 

\begin{align}
	\label{eq:sigmaetaobs}
	\Sigma_{\eta,exp}(\theta^*) &=  \begin{bmatrix} cov_{\lambda_{\eta},l_{\eta}}\left((X_{exp,1},\theta^*),(X_{exp,1},\theta^*) \right) 
		& \cdots 
		&  cov_{\lambda_{\eta},l_{\eta}}\left((X_{exp,1},\theta^*),(X_{exp,N_{exp}},\theta^*) \right) \\
		cov_{\lambda_{\eta},l_{\eta}}\left((X_{exp,2},\theta^*),(X_{exp,1},\theta^*) \right) 
		& \cdots 
		&  cov_{\lambda_{\eta},l_{\eta}}\left((X_{exp,2},\theta^*),(X_{exp,N_{exp}},\theta^*) \right) \\
		\vdots 
		& \ddots & \vdots \\
		cov_{\lambda_{\eta},l_{\eta}}\left((X_{exp,N_{exp}},\theta^*),(X_{exp,1},\theta^*) \right) 
		& \cdots 
		&  cov_{\lambda_{\eta},l_{\eta}}\left((X_{exp,N_{exp}},\theta^*),(X_{exp,N_{exp}},\theta^*) \right) 
	\end{bmatrix}
\end{align}

For full application of the \GP $ \eta_{\lambda_{\eta},l_{\eta}}(x,\theta)$ in the statistical model~(\Eq~\ref{eq:kohGPs}), one further requires the covariance terms between $D_{exp}$ and $D_{sim}$ expressing how close the simulation points are to the individual experimental inputs. This application results in off-diagonal terms $\Sigma_{\eta,exp,sim}$~(\Eq~\ref{eq:sigmaetaobssim}) and  $\Sigma_{\eta,sim,exp} = \Sigma_{\eta,exp,sim}^T$ for the full covariance matrix. 

\begin{align}
	\label{eq:sigmaetaobssim}
	\Sigma_{\eta,exp,sim}(\theta^*) &= \begin{bmatrix} cov_{\lambda_{\eta},l_{\eta}}\left((X_{exp,1},\theta^*),(X_{sim,1},\theta^*) \right) 
		& \cdots 
		&  cov_{\lambda_{\eta},l_{\eta}}\left((X_{exp,1},\theta^*),(X_{sim,N_{sim}},\theta^*) \right) \\
		cov_{\lambda_{\eta},l_{\eta}}\left((X_{exp,2},\theta^*),(X_{sim,1},\theta^*) \right) 
		& \cdots 
		&  cov_{\lambda_{\eta},l_{\eta}}\left((X_{exp,2},\theta^*),(X_{sim,N_{sim}},\theta^*) \right) \\
		\vdots 
		& \ddots & \vdots \\
		cov_{\lambda_{\eta},l_{\eta}}\left((X_{exp,N_{exp}},\theta^*),(X_{sim,1},\theta^*) \right) 
		& \cdots 
		&  cov_{\lambda_{\eta},l_{\eta}}\left((X_{exp,N_{exp}},\theta^*),(X_{sim,N_{sim}},\theta^*) \right) 
	\end{bmatrix}	
\end{align}

\bibliographystyle{tfnlm}

\bibliography{BCKOH_fixed}

\begin{thebibliography}{10}
\providecommand{\url}[1]{\normalfont{#1}}
\providecommand{\urlprefix}{Available from: }

\bibitem{tippingBayesianInferenceIntroduction2004}
Tipping~ME. Bayesian {{Inference}}: {{An Introduction}} to {{Principles}} and
  {{Practice}} in {{Machine Learning}}. In: Bousquet~O, von Luxburg~U,
  Rätsch~G, editors. Advanced {{Lectures}} on {{Machine Learning}}: {{ML
  Summer Schools}} 2003, {{Canberra}}, {{Australia}}, {{February}} 2 - 14,
  2003, {{Tübingen}}, {{Germany}}, {{August}} 4 - 16, 2003, {{Revised
  Lectures}}. Springer; 2004. p. 41--62.
  \urlprefix\url{https://doi.org/10.1007/978-3-540-28650-9_3}.

\bibitem{kennedyBayesianCalibrationComputer2001}
Kennedy~MC, O'Hagan~A. Bayesian calibration of computer models. Journal of the
  Royal Statistical Society: Series B (Statistical Methodology).
  2001;\hspace{0pt}63(3):425--464.
  \urlprefix\url{https://onlinelibrary.wiley.com/doi/abs/10.1111/1467-9868.00294}.

\bibitem{higdonCombiningFieldData2004}
Higdon~D, Kennedy~M, Cavendish~JC, et~al. Combining {{Field Data}} and
  {{Computer Simulations}} for {{Calibration}} and {{Prediction}}. SIAM Journal
  on Scientific Computing. 2004;\hspace{0pt}26(2):448--466.
  \urlprefix\url{https://epubs.siam.org/doi/10.1137/S1064827503426693}.

\bibitem{bayarriComputerModelValidation2007}
Bayarri~MJ, Walsh~D, Berger~JO, et~al. Computer model validation with
  functional output. The Annals of Statistics.
  2007;\hspace{0pt}35(5):1874--1906.
  \urlprefix\url{https://projecteuclid.org/journals/annals-of-statistics/volume-35/issue-5/Computer-model-validation-with-functional-output/10.1214/009053607000000163.full}.

\bibitem{bayarriPredictingVehicleCrashworthiness2009}
Bayarri~MJ, Berger~JO, Kennedy~MC, et~al. Predicting {{Vehicle
  Crashworthiness}}: {{Validation}} of {{Computer Models}} for {{Functional}}
  and {{Hierarchical Data}}. Journal of the American Statistical Association.
  2009;\hspace{0pt}104(487):929--943.
  \urlprefix\url{https://doi.org/10.1198/jasa.2009.ap06623}.

\bibitem{higdonComputerModelCalibration2008}
Higdon~D, Gattiker~J, Williams~B, et~al. Computer {{Model Calibration Using
  High-Dimensional Output}}. Journal of the American Statistical Association.
  2008;\hspace{0pt}103(482):570--583.
  \urlprefix\url{https://www.jstor.org/stable/27640080}.

\bibitem{goldsteinReifiedBayesianModelling2009}
Goldstein~M, Rougier~J. Reified {{Bayesian}} modelling and inference for
  physical systems. Journal of Statistical Planning and Inference.
  2009;\hspace{0pt}139(3):1221--1239.
  \urlprefix\url{https://www.sciencedirect.com/science/article/pii/S0378375808003303}.

\bibitem{qianBayesianHierarchicalModeling2008}
Qian~PZG, Wu~CFJ. Bayesian {{Hierarchical Modeling}} for {{Integrating
  Low-Accuracy}} and {{High-Accuracy Experiments}}. Technometrics.
  2008;\hspace{0pt}50(2):192--204.
  \urlprefix\url{https://www.jstor.org/stable/25471459}.

\bibitem{tuoEfficientCalibrationImperfect2015}
Tuo~R, Wu~CFJ. Efficient calibration for imperfect computer models. The Annals
  of Statistics. 2015;\hspace{0pt}43(6):2331--2352.
  \urlprefix\url{https://projecteuclid.org/journals/annals-of-statistics/volume-43/issue-6/Efficient-calibration-for-imperfect-computer-models/10.1214/15-AOS1314.full}.

\bibitem{plumleeBayesianCalibrationInexact2017}
Plumlee~M. Bayesian {{Calibration}} of {{Inexact Computer Models}}. Journal of
  the American Statistical Association. 2017;\hspace{0pt}112(519):1274--1285.
  \urlprefix\url{https://doi.org/10.1080/01621459.2016.1211016}.

\bibitem{nagelUnifiedFrameworkMultilevel2016}
Nagel~JB, Sudret~B. A unified framework for multilevel uncertainty
  quantification in {{Bayesian}} inverse problems. Probabilistic Engineering
  Mechanics. 2016;\hspace{0pt}43:68--84.
  \urlprefix\url{https://www.sciencedirect.com/science/article/pii/S0266892015300266}.

\bibitem{rasmussenGaussianProcessesMachine2005}
Rasmussen~CE, Williams~CKI. Gaussian {{Processes}} for {{Machine Learning}}.
  The MIT Press; 2005.
  \urlprefix\url{https://direct.mit.edu/books/oa-monograph/2320/Gaussian-Processes-for-Machine-Learning}.

\bibitem{titsiasVariationalLearningInducing2009}
Titsias~M. Variational {{Learning}} of {{Inducing Variables}} in {{Sparse
  Gaussian Processes}}. In: Proceedings of the {{Twelfth International
  Conference}} on {{Artificial Intelligence}} and {{Statistics}}. PMLR; 2009.
  p. 567--574.
  \urlprefix\url{https://proceedings.mlr.press/v5/titsias09a.html}.

\bibitem{burtConvergenceSparseVariational2020}
Burt~DR, Rasmussen~CE, van~der Wilk~M. Convergence of {{Sparse Variational
  Inference}} in {{Gaussian Processes Regression}}. Journal of Machine Learning
  Research. 2020;\hspace{0pt}21(131):1--63.
  \urlprefix\url{http://jmlr.org/papers/v21/19-1015.html}.

\bibitem{niemanAdaptiveSparseVariational2025}
Nieman~D, Szabó~B. Adaptive {{Sparse Variational Approximations}} for
  {{Gaussian Process Regression}}. Bayesian Analysis.
  2025;\hspace{0pt}:1--20\urlprefix\url{https://projecteuclid.org/journals/bayesian-analysis/advance-publication/Adaptive-Sparse-Variational-Approximations-for-Gaussian-Process-Regression/10.1214/26-BA1583.full}.

\bibitem{vecchiaEstimationModelIdentification1988}
Vecchia~AV. Estimation and {{Model Identification}} for {{Continuous Spatial
  Processes}}. Journal of the Royal Statistical Society Series B
  (Methodological). 1988;\hspace{0pt}50(2):297--312.
  \urlprefix\url{https://www.jstor.org/stable/2345768}.

\bibitem{katzfussGeneralFrameworkVecchia2021}
Katzfuss~M, Guinness~J. A {{General Framework}} for {{Vecchia Approximations}}
  of {{Gaussian Processes}}. Statistical Science.
  2021;\hspace{0pt}36(1):124--141.
  \urlprefix\url{https://projecteuclid.org/journals/statistical-science/volume-36/issue-1/A-General-Framework-for-Vecchia-Approximations-of-Gaussian-Processes/10.1214/19-STS755.full}.

\bibitem{williamsUsingNystromMethod2000}
Williams~C, Seeger~M. Using the {{Nyström Method}} to {{Speed Up Kernel
  Machines}}. In: Advances in {{Neural Information Processing Systems}};
  Vol.~13. MIT Press; 2000.
  \urlprefix\url{https://papers.nips.cc/paper_files/paper/2000/hash/19de10adbaa1b2ee13f77f679fa1483a-Abstract.html}.

\bibitem{liuWhenGaussianProcess2020}
Liu~H, Ong~YS, Shen~X, et~al. When {{Gaussian Process Meets Big Data}}: {{A
  Review}} of {{Scalable GPs}}. IEEE Transactions on Neural Networks and
  Learning Systems. 2020;\hspace{0pt}31(11):4405--4423.
  \urlprefix\url{https://ieeexplore.ieee.org/document/8951257}.

\bibitem{trespBayesianCommitteeMachine2000}
Tresp~V. A {{Bayesian Committee Machine}}. Neural Computation.
  2000;\hspace{0pt}12(11):2719--2741.
  \urlprefix\url{https://direct.mit.edu/neco/article/12/11/2719-2741/6426}.

\bibitem{williamsGaussianProcessesRegression1995}
Williams~C, Rasmussen~C. Gaussian {{Processes}} for {{Regression}}. In:
  Advances in {{Neural Information Processing Systems}}; Vol.~8. MIT Press;
  1995.
  \urlprefix\url{https://proceedings.neurips.cc/paper/1995/hash/7cce53cf90577442771720a370c3c723-Abstract.html}.

\bibitem{gelmanPriorDistributionsVariance2006}
Gelman~A. Prior distributions for variance parameters in hierarchical models
  (comment on article by {{Browne}} and {{Draper}}). Bayesian Analysis.
  2006;\hspace{0pt}1(3):515--534.
  \urlprefix\url{https://projecteuclid.org/journals/bayesian-analysis/volume-1/issue-3/Prior-distributions-for-variance-parameters-in-hierarchical-models-comment-on/10.1214/06-BA117A.full}.

\bibitem{lewandowskiGeneratingRandomCorrelation2009}
Lewandowski~D, Kurowicka~D, Joe~H. Generating random correlation matrices based
  on vines and extended onion method. Journal of Multivariate Analysis.
  2009;\hspace{0pt}100(9):1989--2001.
  \urlprefix\url{https://www.sciencedirect.com/science/article/pii/S0047259X09000876}.

\bibitem{hoffmanNoUTurnSamplerAdaptively2014}
Hoffman~MD, Gelman~A. The {{No-U-Turn Sampler}}: {{Adaptively Setting Path
  Lengths}} in {{Hamiltonian Monte Carlo}}. Journal of Machine Learning
  Research. 2014;\hspace{0pt}15(47):1593--1623.
  \urlprefix\url{http://jmlr.org/papers/v15/hoffman14a.html}.

\bibitem{geTuringLanguageFlexible2018}
Ge~H, Xu~K, Ghahramani~Z. Turing: {{A Language}} for {{Flexible Probabilistic
  Inference}}. In: Proceedings of the {{Twenty-First International Conference}}
  on {{Artificial Intelligence}} and {{Statistics}}. PMLR; 2018. p. 1682--1690.
  \urlprefix\url{https://proceedings.mlr.press/v84/ge18b.html}.

\bibitem{xuAdvancedHMCjlRobustModular2020}
Xu~K, Ge~H, Tebbutt~W, et~al. {{AdvancedHMC}}.jl: {{A}} robust, modular and
  efficient implementation of advanced {{HMC}} algorithms. In: Proceedings of
  {{The}} 2nd {{Symposium}} on {{Advances}} in {{Approximate Bayesian
  Inference}}. PMLR; 2020. p. 1--10.
  \urlprefix\url{https://proceedings.mlr.press/v118/xu20a.html}.

\bibitem{smolaSparseGreedyGaussian2000}
Smola~A, Bartlett~P. Sparse {{Greedy Gaussian Process Regression}}. In:
  Advances in {{Neural Information Processing Systems}}; Vol.~13. MIT Press;
  2000.
  \urlprefix\url{https://papers.nips.cc/paper_files/paper/2000/hash/3214a6d842cc69597f9edf26df552e43-Abstract.html}.

\bibitem{lawrenceFastSparseGaussian2002}
Lawrence~N, Seeger~M, Herbrich~R. Fast {{Sparse Gaussian Process Methods}}:
  {{The Informative Vector Machine}}. In: Advances in {{Neural Information
  Processing Systems}}; Vol.~15. MIT Press; 2002.
  \urlprefix\url{https://papers.nips.cc/paper_files/paper/2002/hash/d4dd111a4fd973394238aca5c05bebe3-Abstract.html}.

\bibitem{deisenrothDistributedGaussianProcesses2015}
Deisenroth~MP, Ng~JW. Distributed {{Gaussian}} processes. In: Proceedings of
  the 32nd {{International Conference}} on {{Machine Learning}} - {{Volume}}
  37; ({{ICML}}'15; Vol.~37). JMLR.org; 2015. p. 1481--1490.

\bibitem{csatoSparseOnLineGaussian2002}
Csató~L, Opper~M. Sparse {{On-Line Gaussian Processes}}. Neural Computation.
  2002;\hspace{0pt}14(3):641--668.
  \urlprefix\url{https://doi.org/10.1162/089976602317250933}.

\bibitem{liuGeneralizedRobustBayesian2018b}
Liu~H, Cai~J, Wang~Y, et~al. Generalized {{Robust Bayesian Committee Machine}}
  for {{Large-scale Gaussian Process Regression}}. In: Proceedings of the 35th
  {{International Conference}} on {{Machine Learning}}. PMLR; 2018. p.
  3131--3140. \urlprefix\url{https://proceedings.mlr.press/v80/liu18a.html}.

\bibitem{willowSparseBayesianCommittee2025}
Willow~SY, Kim~S, Yang~DC, et~al. A sparse {{Bayesian Committee Machine}}
  potential for oxygen-containing organic compounds. Chemical Physics Reviews.
  2025;\hspace{0pt}6(2).
  \urlprefix\url{https://pubs.aip.org/aip/cpr/article/6/2/021401/3344282/A-sparse-Bayesian-Committee-Machine-potential-for}.

\bibitem{coxStatisticalMethodTuning2001}
Cox~DD, Park~JS, Singer~CE. A statistical method for tuning a computer code to
  a data base. Computational Statistics \& Data Analysis.
  2001;\hspace{0pt}37(1):77--92.
  \urlprefix\url{https://www.sciencedirect.com/science/article/pii/S0167947300000578}.

\bibitem{eldredInvestigationReliabilityMethod2007}
Eldred~MS, Agarwal~H, Perez~VM, et~al. Investigation of reliability method
  formulations in {{DAKOTA}}/{{UQ}}. Structure and Infrastructure Engineering.
  2007;\hspace{0pt}3(3):199--213.
  \urlprefix\url{https://doi.org/10.1080/15732470500254618}.

\bibitem{condeResearchProgramRecent2017}
Conde~M, Antipov~S, Doran~D, et~al. Research {{Program}} and {{Recent Results}}
  at the {{Argonne Wakefield Accelerator Facility}} ({{AWA}}). JACOW, Geneva,
  Switzerland; 2017. p. 2885--2887.
  \urlprefix\url{https://proceedings.jacow.org/ipac2017/doi/JACoW-IPAC2017-WEPAB132.html}.

\bibitem{adelmannOPALVersatileTool2019}
Adelmann~A, Calvo~P, Frey~M, et~al. {{OPAL}} a {{Versatile Tool}} for {{Charged
  Particle Accelerator Simulations}}; 2019.
  \urlprefix\url{http://arxiv.org/abs/1905.06654}.

\end{thebibliography}

\end{document}